\documentclass[12pt]{article}

\usepackage[english]{babel}
\usepackage{graphicx}
\usepackage{epstopdf}
\usepackage{color}
\usepackage{ragged2e}
\usepackage{float}
\usepackage{xcolor}
\usepackage{times}
\usepackage{authblk}
\usepackage[final]{pdfpages}

\title{Local Sensing of Correlated Electrons in Dual-moir\'e Heterostructures using Dipolar Excitons}

\date{}
\author[1]{Weijie Li}
\author[1]{Luka M. Devenica}
\author[2]{Jin Zhang}
\author[3]{Yang Zhang}
\author[4]{Xin Lu}
\author[5]{Kenji Watanabe}
\author[5]{Takashi Taniguchi}
\author[2,6,7]{Angel Rubio}
\author[1]{Ajit Srivastava$^{\ast}$}
\affil[1]{Department of Physics, Emory University, 30322 Atlanta, Georgia, USA}
\affil[2]{Center for Free Electron Laser Science, Max Planck Institute for the Structure and Dynamics of Matter, 22761 Hamburg, Germany}
\affil[3]{Department of Physics, Massachusetts Institute of Technology, 02139 Cambridge, Massachusetts, USA}
\affil[4]{Department of Physics, Tulane University, 70118 New Orleans, Louisiana, USA}
\affil[5]{National Institute for Materials Science, Namiki 1-1, Tsukuba, Ibaraki 305-0044, Japan}
\affil[6]{Center for Computational Quantum Physics, Simons Foundation Flatiron Institute, 10010 New York, New York, USA}
\affil[7]{Nano-BioSpectroscopy Group, Departamento de Fisica de Materiales,Universidad del Pa{\'i}s Vasco, 20018 San Sebastián, Spain}

\begin{document} 


\baselineskip24pt


\maketitle 
\normalsize{$^\ast$To whom correspondence should be addressed; E-mail:  ajit.srivastava@emory.edu}


{\bf Moir\'e heterostructures are rapidly emerging as a tunable platform to study correlated electronic phenomena. Discovery of exotic quantum phases in moir\'e systems requires novel probes of charge and spin order. Unlike detection schemes which average over several moir\'e cells, local sensors can provide richer information with greater sensitivity. We study a WSe$_2$/MoSe$_2$/WSe$_2$ heterotrilayer which hosts excitons and electrons in distinct moir\'e lattices, and show that localized dipolar excitons are sensitive proximity charge sensors, uncovering numerous correlated electronic states at fractional fillings of the multi-orbital moir\'e lattice. In addition, the emission polarization can reveal the local electronic spin configuration at different fillings. Our results establish dipolar excitons as promising candidates to study emergent quantum matter and quantum magnetism in moir\'e crystals with higher spatial resolution.}

When the strength of Coulomb interactions between electrons in a solid dominates their kinetic energy, electrons often display correlated behavior. Such correlations can lead to fascinating new phases of matter beyond the traditional paradigm, whose pursuit is one of the main themes of modern condensed matter physics. Moir\'e heterostructures, formed by introducing a small twist between vertically stacked two-dimensional (2D) crystals, offer an easy knob to tune electronic correlations by dramatically reducing the electronic kinetic energy~\cite{bistritzer2011moire, lisi2021observation}. Moreover, the ability to control moir\'e lattice structure and in-situ tuning of interactions makes this system promising for solid-state quantum simulation and systematic exploration of rich electronic phase diagrams. Recently, several quantum phases with charge and spin ordering such as Wigner crystals\cite{regan2020mott,zhou2021bilayer,smolenski2021signatures}, Mott insulators\cite{shimazaki2021optical} and quantum anomalous Hall states\cite{li2021quantum, chen2020tunable, polshyn2020electrical} have been observed in moir\'e heterostructures of graphene and semiconducting transition metal dichalcogenides (TMDs), with predictions for several other undiscovered exotic phases~\cite{kennes2021moire}.
%


The experimental detection of such phases in semiconducting TMD moir\'e heterostructures is an ongoing endeavor which presents challenges but also unique opportunities. For example, while electronic transport measurements are plagued by high electrical contact resistance~\cite{allain2015electrical, li2021charge}, inherently strong light-matter interactions have been successfully exploited in non-contact optical spectroscopic techniques to uncover correlated insulating states at several fractional fillings of the moir\'e lattice~\cite{ shimazaki2021optical, li2021charge, tang2020simulation, xu2020correlated, jin2021stripe, liu2021excitonic, huang2021correlated}. One of the most attractive features of van der Waals (vdW) materials is that an additional 2D material serving as a proximity sensor of the correlated electronic behavior can be easily incorporated in the same heterostructure during the fabrication process~\cite{xu2020correlated, li2021imaging}. However, the existing detection techniques probe over tens to several thousand moir\'e unit cells and lack the spatial resolution needed to study local fluctuations of charge and spin order. Besides traditional scanning probe techniques for local sensing~\cite{li2021imaging}, vdW materials offer an all-2D approach wherein localized charge and spin sensors in one moir\'e lattice spatially sample correlated electronic states in a different moir\'e lattice of the same heterostructure, from which a global picture can be reconstructed.


Taking a first step towards this approach, we study a dual-moir\'e WSe$_2$/MoSe$_2$/WSe$_2$ heterotrilayer which features different moir\'e superlattices for electrons and dipolar interlayer excitons, owing to different twist angles of the top and bottom heterobilayers~\cite{tong2020interferences}. Moir\'e-trapped interlayer excitons (IX) with stable and spectrally narrow emission, together with an out-of-plane, non-oscillating dipole, respond sensitively to electric fields by changing their emission energy, thus becoming local sensors of charge distribution around them. Moreover, the degree of circular polarization (DCP) of their emission is sensitive to tiny magnetic fields and in turn to spin phenomena. Crucially, the excitonic moir\'e lattice hosting local sensors is distinct from the electronic moir\'e, which enables efficient sampling of symmetry-broken correlated electronic states in the MoSe$_2$ layer as the electron density is increased. An additional feature of such a dual-moir\'e heterostructure is that the electronic moir\'e lattice can exhibit multiple degenerate minima due to interference of the two moir\'e potentials, thus realizing multi-orbital Hubbard models with richer physics~\cite{florens2002mott, zhang2021electronic}. Our observations add highly sensitive, moir\'e-based local probes to existing optical techniques by uncovering correlated electronic states in a dual-moir\'e heterostructure.

Figure 1A shows an image of our device consisting of hBN encapsulated WSe$_2$/MoSe$_2$/WSe$_2$ heterotrilayer with dual graphite gates which allow for independent control of electron density and out-of-plane displacement field (Fig.~1B)~\cite{methods}. The type-II band alignment in MoSe$_2$/WSe$_2$ heterobilayer results in the lowest energy state for electrons (holes) in the MoSe$_2$ (WSe$_2$) layer such that the top (bottom) IX formed in the top (bottom) heterobilayer has a dipole moment pointing up (down), which responds to an out-of-plane electric field ($E$). The effect of different top and bottom twist angles is shown in Fig.~1C, which plots the calculated moir\'e potentials for an IX in the bottom heterobilayer and an electron in the MoSe$_2$ layer for twist angles of 1$^\circ$/4$^\circ$, representative of our sample. While an IX experiences only the bottom moir\'e potential, an electron sees the interference of the top and bottom moir\'e potentials resulting in a multi-orbital (multi-minima) electron moir\'e potential~\cite{tong2020interferences}. 

Low temperature photoluminescence (PL) spectra of the sample show several localized emitters in the IX energy range, which are non-jittering with extremely narrow, instrument-limited linewidth of $\sim$ 26 $\mu$eV (Fig.~1, D and E). We identify the two species of IX with opposite dipoles by their opposite Stark shifts under $E$ (Fig.~1F). Their circularly polarized emission and Land\'e g-factor value indicates them to be moir\'e trapped IXs~\cite{seyler2019signatures} (see Fig.~S1, S2). As bottom IXs are abundant in our sample, in the following we focus primarily on them. In addition to sensing electric fields through changes in their energy, IX can sense magnetic fields ($B$) as small as 10 mT. As shown in Fig.~1G, DCP under circularly polarized excitation rapidly increases from zero to unity with $B$. This effect can be used to sense local spin or valley configuration of electrons and enable probing of quantum magnetism. Figure 1H shows a cartoon depiction of our local charge sensing scheme -- strong electron-electron interactions (U$_{e-e}$) result in correlated, charge-ordered states at fractional fillings of the electronic moir\'e lattice as carrier density is changed, which are then sensed through energy shifts of localized IX arising from electron-dipole interactions (U$_{e-d}$). IXs localized at different excitonic-moir\'e sites sample the correlated electronic state efficiently due to distinct moir\'e lattices for electrons and IX. 

As we increase the electron density in the sample under $E$ = 0, stable and sharp emission of localized IX (Fig.~1D) starts exhibiting several seemingly random spectral jumps, however, upon reversing the gate voltage (V$_\mathrm{g}$), even the minutest spectral jumps ($\sim$ 50 $\mu$eV) are remarkably reproduced (Fig.~2A, Fig.~S3, Fig.~S4). We have performed V$_\mathrm{g}$ scans over a period of several months and the spectral jumps are perfectly reproduced. Thus, we can conclude that spectral jumps of IX, caused by addition of electrons to the sample, are not random but highly deterministic in V$_\mathrm{g}$ and arise due to the change in charge configuration near localized IXs. In other words, electrons are being added to a potential landscape which is static and can be reproducibly populated with V$_\mathrm{g}$.

To reveal any correlations in the jump behavior and signatures of order in the potential landscape, we perform simultaneous V$_\mathrm{g}$ scans over two different regions (spot A and spot B) of the sample which are separated by $\sim$ 4 $\mu$m, several times the excitation spot size. As shown in Fig.~2B, after starting with stable emission, we find that the PL qualitatively changes in both spots at a V$_\mathrm{g}$ marked by the dotted line L$_\mathrm{I}$, followed by a sudden redshift of about 6-7 meV for several IXs in both regions at a V$_\mathrm{g}$ marked by the dotted line L$_\mathrm{T}$. L$_\mathrm{I}$ marks the end of intrinsic region whereas the redshift at L$_\mathrm{T}$ is consistent with the recently observed moir\'e trion\cite{wang2021moire,liu2021signatures, PhysRevX.11.031033} where the excess electron resides in the same moir\'e unit cell as the IX (See supplementary Fig.~ S5 materials for further data on moir\'e trions). At higher electron densities, PL from both regions broadens and becomes weak. Besides these global features, we also find several V$_g$ values where spectral jumps occur simultaneously, indicated by dashed white lines and dashed red ellipses in Fig.~2B, both within the same spot and across the two spots. The bottom panels of Fig.~2B show other such correlated jumps with non-zero $E$ which helps spread the jump in gate voltage range for easier identification (see Fig.~S6, S7, S10 for further correlations in jump behavior). More importantly, we observe that in addition to an overall blue shift, jumps occur as both red and blue shifts, which is inconsistent with a simple picture where a monotonic increase of the electronic density in the middle MoSe$_2$ layer only increases the emission energy of IX~\cite{baek2021coulombstaircase}. In the following, we analyze these correlations in greater detail.

Figure 2C shows the gate-dependent PL behavior of localized IX together with the global intralayer exciton reflectance from the excitation spot size. We identify 1s and 2s excitons of MoSe$_2$ and WSe$_2$ intralayer excitons which allows us to identify the intrinsic region marked by top red dashed line, labeled as $\nu^*$ = 0. Extending this line to the PL scan, as expected, we find that it lies in the stable region. As we are focusing on the electron-doped side, we identify a V$_g$ labeled as $\nu^*$ = 1 where the reflectance of both Mo and W excitons first develops a kink. The terminology for this assignment will become clear later. The PL also develops a kink at this gate voltage and with higher doping, both reflectance and PL get broader and eventually disappear together (Fig. S11, S12, S13 show that at the PL data also shows kinks which correlate with the global reflectance). Thus, while reflectance is not sensitive enough to capture the numerous jumps observed in the PL of localized IXs, we do find that both PL and reflectance show several global features which are qualitatively similar. 


Next, we replot the data in Fig.~2B with the V$_g$ axis relabeled in fractions of the voltage interval defined by $\nu^* = 0$ and $\nu^* = 1$ (Fig.~3, A and B). Remarkably, this simple relabeling of axis shows that several jumps for both spots occur at gate voltages corresponding to integer multiples of $\nu^*$ = 1/12. We note that our choice of 1/12 as the minimum fraction is much larger than the step size of gate voltage scan which is $\sim$1/400, thus ruling out accidental correlations in jumps. As shown in Fig.~3A, the jumps which do not fall on 1/12 graduations (marked by dashed white lines) are often separated by other such jumps by a voltage range equal to 1/3 or 1 in units of $\nu^*=1$. To further confirm that this trend is statistically significant, we plot occurrences of $\nu^*$ modulo 1/12 of over 330 jumps in a histogram. If most of the jumps occur at integer multiples of 1/12, then we should observe a peak at zero. As shown in Fig.~3C, there is indeed such a peak for both spots A and B. Furthermore, the residual peaks seem to be centered at 1/3 and 1/2 of the 1/12 graduation. The probability to obtain the observed peak value at zero from a completely random distribution of jumps is estimated to be $\sim$5 \% ($\sim$10\%) for spot A (B).  This probability will reduce further if we include the residual peaks at 1/3 and 1/2. Thus, we conclude that the spectral jumps of IX occur at certain rational fillings of the electronic moir\'e lattice which are in correspondence with $\nu^*$. 

While electrons are added continuously, we observe jumps primarily at fractional fillings which can be understood to arise from the sudden, global (over 4 $\mu$m) redistributions of electrons at these fillings, which are locally sensed by moir\'e-trapped dipolar IXs or moir\'e trions. The energy of spectral jumps is expected to be the same for a dipolar IX or moir\'e trion because the loosely bound excess electron is present in both the initial and final states of the emission process. The redistribution of charges at certain fractional fillings also explains the observed red shifts of localized IXs. This constitutes strong evidence for incompressible electronic crystalline states with interaction-induced, long range order at fractional fillings of heterotrilayer moir\'e potential. The transitions between these charge-ordered correlated states are sensed by localized IX located at different positions with respect to the electronic moir\'e unit cell. The abundance of fractional fillings which are multiples of 1/12 is expected because of the three-fold symmetry and reconstruction of periodicity which gives quarter fillings on the underlying moir\'e lattice. Moreover, choosing other fractions such as 1/11 and 1/13 do not yield peaks at zero of the histogram (Fig.~S14). Finally, we note that the overall blueshift with higher doping is expected even in the absence of correlated states.

In order to relate $\nu^*$ to the actual filling fraction $\nu$, we first note that defining $\nu$ (or even a moir\'e length) for a multi-orbital moir\'e lattice is ambiguous. Nonetheless, we estimate the doping density corresponding to $\nu^*$ = 1 from a capacitor model to be 1.7$\times$10$^{12}$ cm$^{-2}$ and upon setting $\nu^*$ = $\nu$ = 1, we obtain an effective moir\'e length of $\sim$ 7-8 nm (see Supplementary text). This length scale is the effective electron lattice constant, which is smaller than the 1$^\circ$ moir\'e length, implying that not only first orbital is involved in the doping behavior for the 1$^\circ$/4$^\circ$ case. In order to confirm the salient features of our observations and the explanation provided in the previous paragraph, we simulate the jump behavior of IX trapped in a moir\'e potential (Fig.~1C, left panel) by calculating their energies at various symmetry-broken electronic crystalline states of the heterotrilayer moir\'e lattice (Fig.~1C, right panel). Two such electronic crystals simulated by classical simulated annealing technique (see Supplementary text) are shown in Fig.~3D at different fillings, $n_\mathrm{e}$, of the multi-orbital moir\'e lattice in Fig.~1C. Figure 3E shows the simulation of energy shifts for a representative excitonic moir\'e site as a function of $n_\mathrm{e}$. Indeed, we observe both red and blue shifts with an overall blue shift, in good agreement with our data and interpretation. In addition to a qualitative comparison, we can perform statistics of the red and blue shifts for spots A and B, as shown in Fig.~3F (left panel), which also agrees very well with the spectral jump statistics of our simulation (Fig.~3F, right panel). 

Finally, we convert the size of spectral jumps into an effective length using a charge sensing model in which the energy shift of IX arises due to the electric field of an electron at a distance $r_\mathrm{eff}$ (see Fig.~S15). Figure 3G shows that $r_\mathrm{eff}$ is peaked at 7-8 nm for both data and simulation, in excellent agreement with our independent estimate of an effective moir\'e length obtained from the capacitor model. Owing to an uncertainty in the twist angle of $<$ 1$^\circ$ between MoSe$_2$ and WSe$_2$ layer, we also simulate the jump behavior for a heterotrilayer for twist angles with 0$^\circ$/3$^\circ$. In this case, the dual-moir\'e potential arises from strain relaxation which is captured by DFT calculations and shows a multi-orbital moir\'e unlike a heterobilayer with a twist of 3$^\circ$ (see Supplementary text, Fig.~S16). Fig.~S17 shows that there is qualitatively similar behavior of jumps even in this case. Comparing our data to simulations for multi-orbital moir\'e lattices, we find evidence that a different orbital is occupied after half-filling of the first moir\'e minima, which is expected because of the large on-site repulsion (see Supplementary Fig.~S17, S18).

Figure 4 shows the DCP of PL as a function of electron density. We note that most localized IX have negligible DCP in the intrinsic region below 10 mT, very likely due to a residual electron-hole (e-h) exchange interaction, $J_\mathrm{eh}$ which mixes the K and -K valleys~\cite{yu2014dirac}. As electrons are added, the DCP jumps to a finite value once moir\'e trions are created. This can be explained by the singlet configuration of the two electrons in moir\'e trion, which quenches $J_\mathrm{eh}$~\cite{lu2019optical}. At the highest doping before which the emission broadens and disappears, DCP remains large. However, for certain moir\'e trions, we see a non-monotonic behavior of DCP with doping. Dashed boxes in Fig.~4 A to C show that the DCP suddenly vanishes at certain filling while the PL is still strong and recovers at a later fractional filling. Moreover, like the neutral IX, the negligible DCP in this intermediate electron density becomes finite with a tiny magnetic field of  50 mT\cite{smolenski2016tuning}, as shown in Fig.~4D (see Supplementary Fig.S19, S20 for more data).


To explain the disappearance of DCP at intermediate doping, we invoke a resonant tunneling-induced recovery of e-h exchange as shown in Fig.~4E. At intermediate filling $\nu$, 0 $<$ $\nu$ $<$ 1, when a moir\'e site spatially close to the moir\'e trion is singly occupied with its energy close to the electrons comprising the trion, there can be strong resonant tunneling ($t$) resulting in an effective e-h exchange interaction (Fig.~4E). This process requires spin-valley conserving tunnelings of trion electrons and the crystal electron together with $J_\mathrm{eh}$ and is equivalent to an effective spin flip-flop process between the trion hole and a crystal electron which gives a tunnel-dependent exchange of $\sim$ $t^2 J_\mathrm{eh}/\delta^2$ (See supplementary materials text). Such a process requires sufficient proximity of the electron to the moir\'e trion, a condition which may not be satisfied by all localized IX. Although this process is second order in $t$, it can become large at resonance. Crucially, this process requires half-filling of the participating moir\'e site to enable spin-flipping of the electron. At higher doping densities, when the moir\'e site has double occupancy, this resonant tunneling-induced exchange is either Pauli blocked or the resonance condition is lost due to additional repulsion energy and the DCP recovers, consistent with our observations that the recovery of DCP occurs above $\nu > 1$. Thus, the DCP of IX gives local spin information of correlated electronic states and can be exploited for probing quantum magnetism in the future.

In conclusion, we have introduced a moir\'e-based local sensing scheme for correlated electronic behavior in a monolithic dual-moir\'e heterostructure, relying on sampling rather than scanning of the sample. Data from several localized sensors, each probing $\sim$ 10 nm, is used to uncover long-ranged charge-ordered electronic states as a function of electron density. In addition, the spin-valley degree of dipolar sensors yields information about the local spin behavior which could be used to detect magnetic order in the future. Our scheme, also based on optical spectroscopy, complements the recently employed techniques such as reflectance spectroscopy, and is potentially more sensitive by not averaging over disorder or local domains. Our multi-orbital moir\'e structure could be used to simulate multi-orbital Hubbard model and explore quantum magnetism with Hund's coupling~\cite{georges2013strong}. In addition to vdW heterostructures, proximity sensing using 2D dipolar excitons could be used to study other strongly correlated materials such as unconventional superconductors and quantum magnets.

Our data contains a wealth of local information about correlated electronic states and in future can be analyzed using machine learning techniques to reconstruct a global picture. In addition to static structure, it should be possible to study dynamics from local charge and spin fluctuations, which are expected to be enhanced near quantum phase transitions. For example, the quantum nature of dipolar emitters can be exploited in photon statistics measurements which could be affected by such fluctuations. Finally, in addition to being passive sensors, localized dipolar IXs can strongly couple to other many-body systems and act as quantum impurities with an optical readout.

\justify
{\bf Acknowledgments}  We acknowledge many enlightening discussions with Martin Claassen, Benjamin Feldman, Mohammad Hafezi, Ata\c{c} Imamo\u{g}lu, Tomasz Smole\'nski, Alexander Popert, Patrick Kn\"uppel and Lede Xian. 
\justify
{\bf Funding}  This work was supported by the EFRI program-grant (\# EFMA-1741691 for A. S.) and NSF DMR award (\# 1905809 for A. S. ). The theoretical work was supported by the European Research Council (ERC-2015-AdG694097), cluster of Excellence AIM, SFB925 and Grupos Consolidados (IT1249-19). We acknowledge support by the Max Planck Institute-New York City Center for Non-Equilibrium Quantum Phenomena. The Flatiron Institute is a division of the Simons Foundation. J.Z. acknowledges funding received from the European Union Horizon 2020 research and innovation program under Marie Sklodowska-Curie Grant Agreement 886291 (PeSD-NeSL).  
\justify
{\bf Author contributions}
 A. S., W. L., X. L. conceived the project. K. W., T. T. provided the hBN crystal. W. L. prepared the samples. W. L., L. D. carried out the measurements. J. Z., Y. Z. conducted the DFT calculations. W. L. conducted the simulation. A. S., A. R. supervised the project. All authors were involved in analysis of the experimental data and contributed extensively to this work.


\clearpage

\begin{figure}
\includegraphics[scale=0.9]{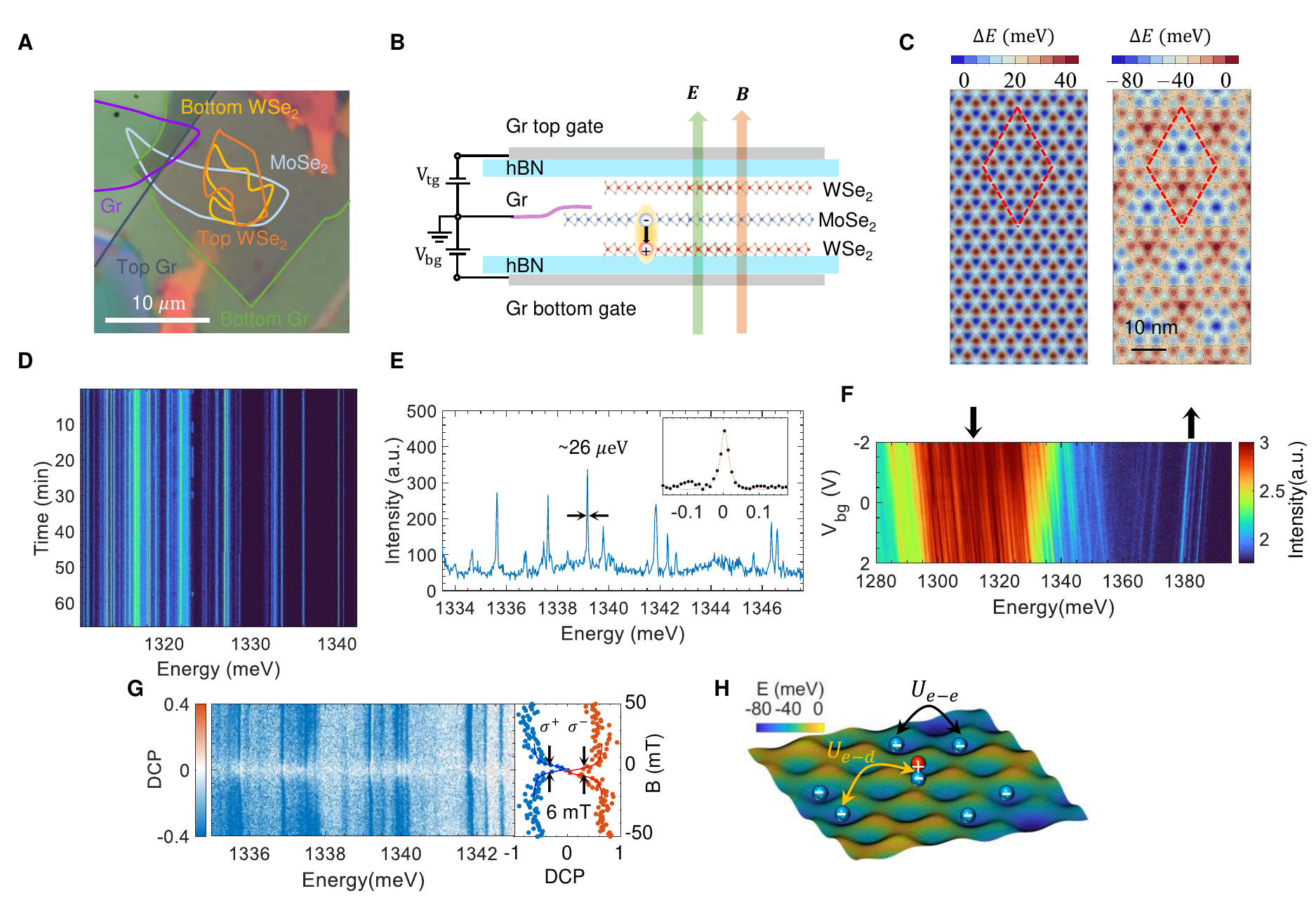}
\end{figure}
\noindent {\bf Fig. 1. Sharp, non-jittering localized dipolar excitons in WSe$_2$/MoSe$_2$/WSe$_2$ heterostructures as electric and magnetic field sensors.} {\bf(A)} Optical microscope image and {\bf(B)} Side-view illustration of the dual-gate WSe$_2$/MoSe$_2$/WSe$_2$ heterostructure. The bottom WSe$_2$ layer is rotated by 3$^\circ$ relative to the top WSe$_2$ layer. {\bf(C)} The calculated bottom exciton potential (left) and electron potential in the middle MoSe$_2$ layer (right) in a 1$^\circ$/4$^\circ$ trilayer heterostructure~\cite{tong2020interferences}. The electron moir\'e unit cell is outlined by the red dashed lines. {\bf(D)} The time-trace photoluminescence (PL) emission of localized interlayer excitons (IXs), showing sharp and stable peaks. The energies of the excitons are consistent with IX energies. {\bf(E)} The PL spectra of IXs showing instrument resolution-limited linewidths as narrow as $\sim$ 26 $\mu$eV. {\bf(F)} Gate dependence of dipolar IX emission under applied  voltage to the back gate with top gate and trilayer grounded. The electric field from the asymmetric gating induces blueshift (redshift) of the IX in the bottom (top) heterobilayer denoted by downward (upward) from -2 V to 2 V. The black arrows represent the dipole directions of the corresponding excitons. The intensity colorbar is logarithmic. {\bf(G)} Polarization-resolved magneto-PL of IXs with $\sigma^+$ excitation. The degree of circular polarization (DCP) increases with both positive and negative magnetic fields {\it B}. The Lorentzian fitting of the DCP versus {\it B} gives a 6 mT width. {\bf(H)} Schematic of our charge-order sensing scheme using localized IXs. The Coulomb interaction between electrons ($U_{e-e}$) with the moir\'e potential gives rise to electronic crystallization, while the IXs can sense the electrons by the Coulomb interaction between electrons and dipoles ($U_{e-d}$). The excitation laser is at 1.70 eV and 50 nW for (D)(E)(G) and 600 nW for (F).

\clearpage

\begin{figure}
\includegraphics[scale=0.9]{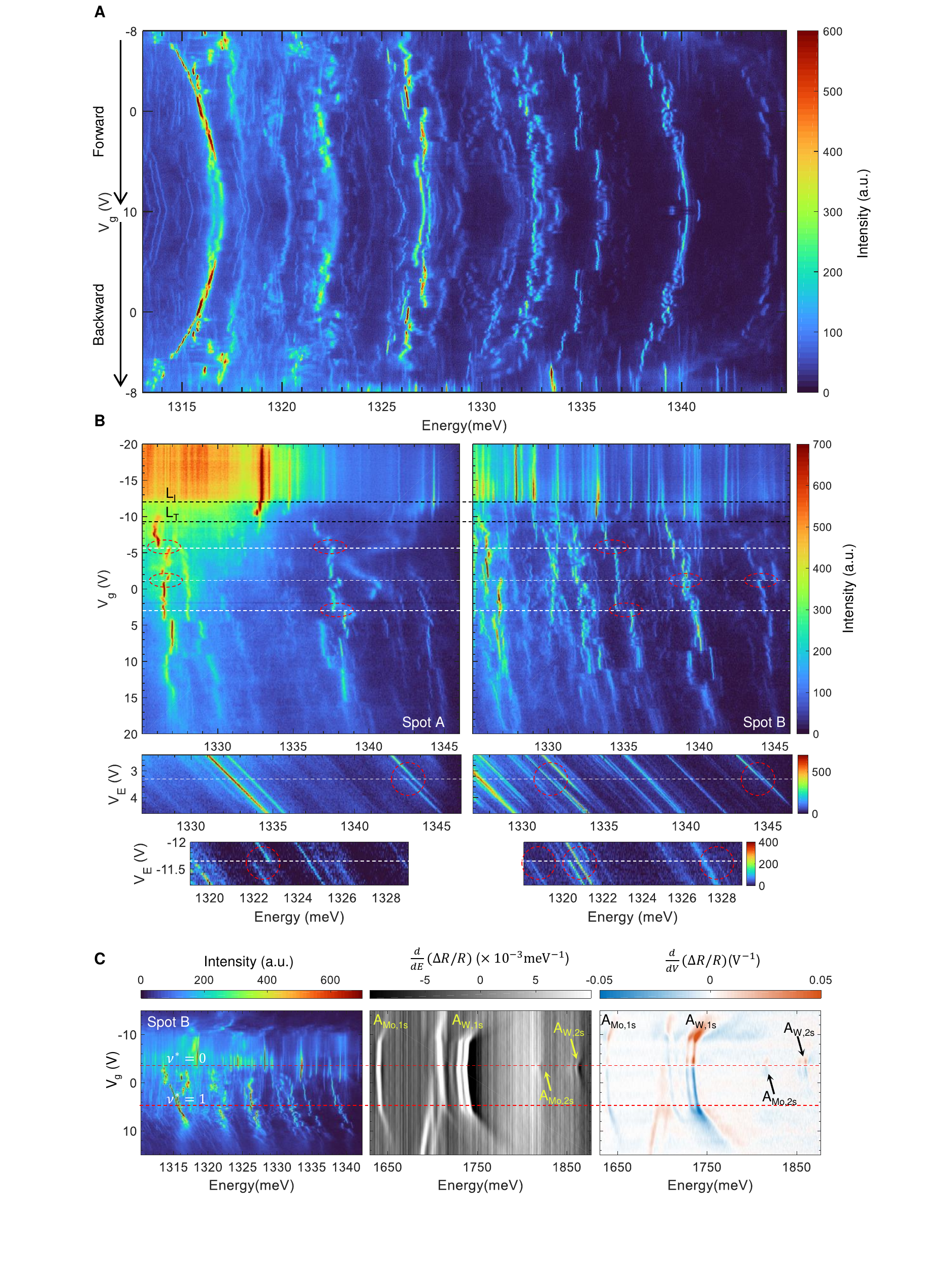}
\end{figure}
\clearpage

\noindent {\bf Fig. 2. Features of correlated electrons in the doping-dependent energy shifts of IXs.} {\bf(A)} Reproducible red- and blue-shifts of IX energies with electron doping. The symmetric gating V$_\mathrm{g}$ from -8 to 10 V (forward) introduces electrons into the system without electric field. Sudden spectral jumps of IX energies and overall blue-shift are caused by changes in Coulomb interactions between electrons and dipoles ($U_{e-d}$). The spectral jumps are remarkably-well reproduced when V$_\mathrm{g}$ is reversed from 10 V to -8 V (backward), excluding the possibility of random jumps. {\bf(B)} Correlation of energy shifts between different IXs and different sample positions. Simultaneous PL gate scan of spot A (left) and spot B (right), which are separated by 4 $\mu$m, shows similar global features: no energy shift above line L$\mathrm{_I}$, appearance of red-shifted peaks after line L$\mathrm{_T}$ and broadening of peaks after 10 V. The spectral jumps occur at the same voltages (white dashed lines) for different IXs at the same spot and across different spots, as shown by the energy shifts outlined by the red dashed ellipses. The bottom four PL gate scans with asymmetric gating show similar correlated energy shifts, now with both electric field and doping serving to stretch and therefore increasing the resolution of each energy shift. {\bf(C)} Correlation between doping-dependent PL (left) and reflectance contrast spectra (middle and rights). A$\mathrm{_{Mo,1s}}$, A$\mathrm{_{W,1s}}$, A$\mathrm{_{Mo,2s}}$ and A$\mathrm{_{W,2s}}$ are the MoSe$_2$ and WSe$_2$ intralayer 1s and 2s exciton resonances, respectively. The first red dashed line, which is at the charge neutral point, is assigned as $\nu^*$ = 0. The second red dashed line, where the reflectance shows a kink, is assigned as $\nu^*$ = 1. The voltage range difference between PL in (A) and (C) is because of change in gating behavior after subjecting the sample to high voltage ($\sim$ 40 V) and is corrected for (see supplementary text, Fig.~S8, S9). The excitation laser is at 1.70 eV and 80 nW.

\clearpage

\begin{figure}
\includegraphics[scale=0.9]{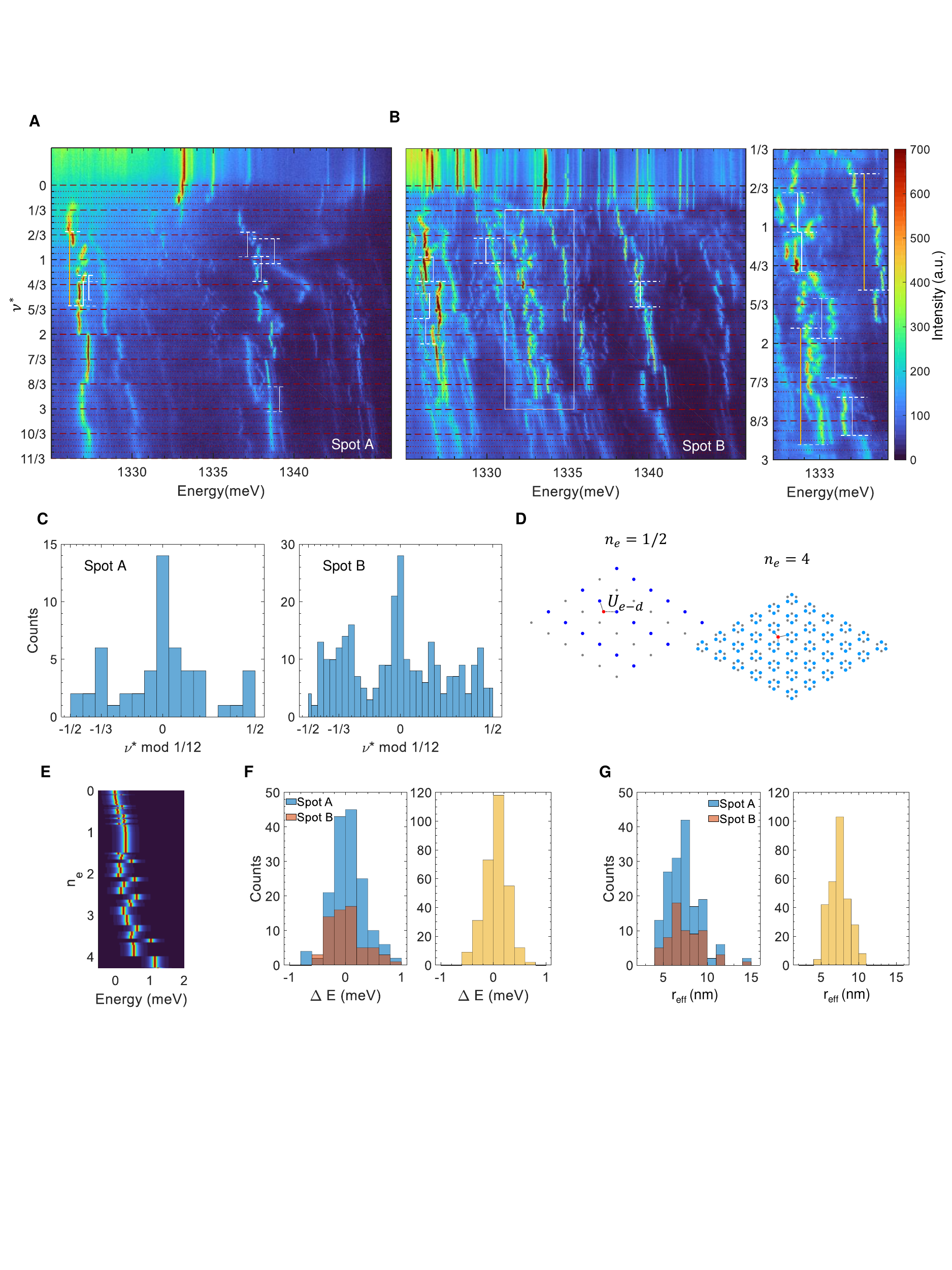}
\end{figure}
\clearpage

\noindent {\bf Fig. 3. Filling-fraction assignment and simulation of the correlated electronic states.} {\bf(A)} Doping dependence of PL emission at spot A with respect to the filling fractions $\nu^*$. The dashed dark red lines indicate fractional fillings of the electronic moir\'e superlattice at multiples of $\nu^*$ = 1/12. Most of the spectral jumps fall on these lines. The horizontal white dashed lines denote the energy shifts not falling on 1/12 fractions, however, they are often separated from each other by intervals corresponding to $\Delta\nu^*$ = 1/3 (vertical white solid line) or $\Delta\nu^*$ = 1 (vertical yellow solid line). {\bf(B)} Doping dependence of PL emission at spot B with respect to $\nu^*$ (left) with zoomed-in view for the white box (right). {\bf(C)} Histograms of $\nu^*$ corresponding to spectral jumps in spot A (left) and spot B (right) modulo 1/12. The histograms peak at 0, confirming that jumps preferentially occur at fillings which are integer multiples of 1/12. {\bf(D)} Charge-ordered states from classical simulated annealing simulation in 1$^\circ$/4$^\circ$ electron potential for filling factors $n_e$ = 1/2 and $n_e$ = 4, where $n_e$ is the number of electrons per unit cell with a periodicity of 18 nm. The deep (light) blue dots in the left (right) panel are the first (second) orbital sites occupied by electrons and grey dots are unoccupied sites. The localized dipolar excitons (red dots) sense electron-dipole interaction ($U_{e-d}$) from occupied electrons. {\bf(E)} Calculated energy of the localized dipolar exciton show red- and blue-shifts and an overall blueshift, consistent with experimental features. {\bf(F)} Histograms of spectral jump size $\Delta E$ from the experiment (left) and simulation (right). {\bf(G)} Histograms of effective length scale for the experiment (left) and simulation (right) obtained by a procedure described in Supplementary text and Fig.~S15. Both experimental and simulated results give an effective length of 7-8 nm. The $\nu^*$ where spectral jumps occur and jump size ($\mathrm{\Delta E}$) are manually determined for all distinguishable IX peaks.  
\clearpage

\begin{figure}
\includegraphics[scale=0.9]{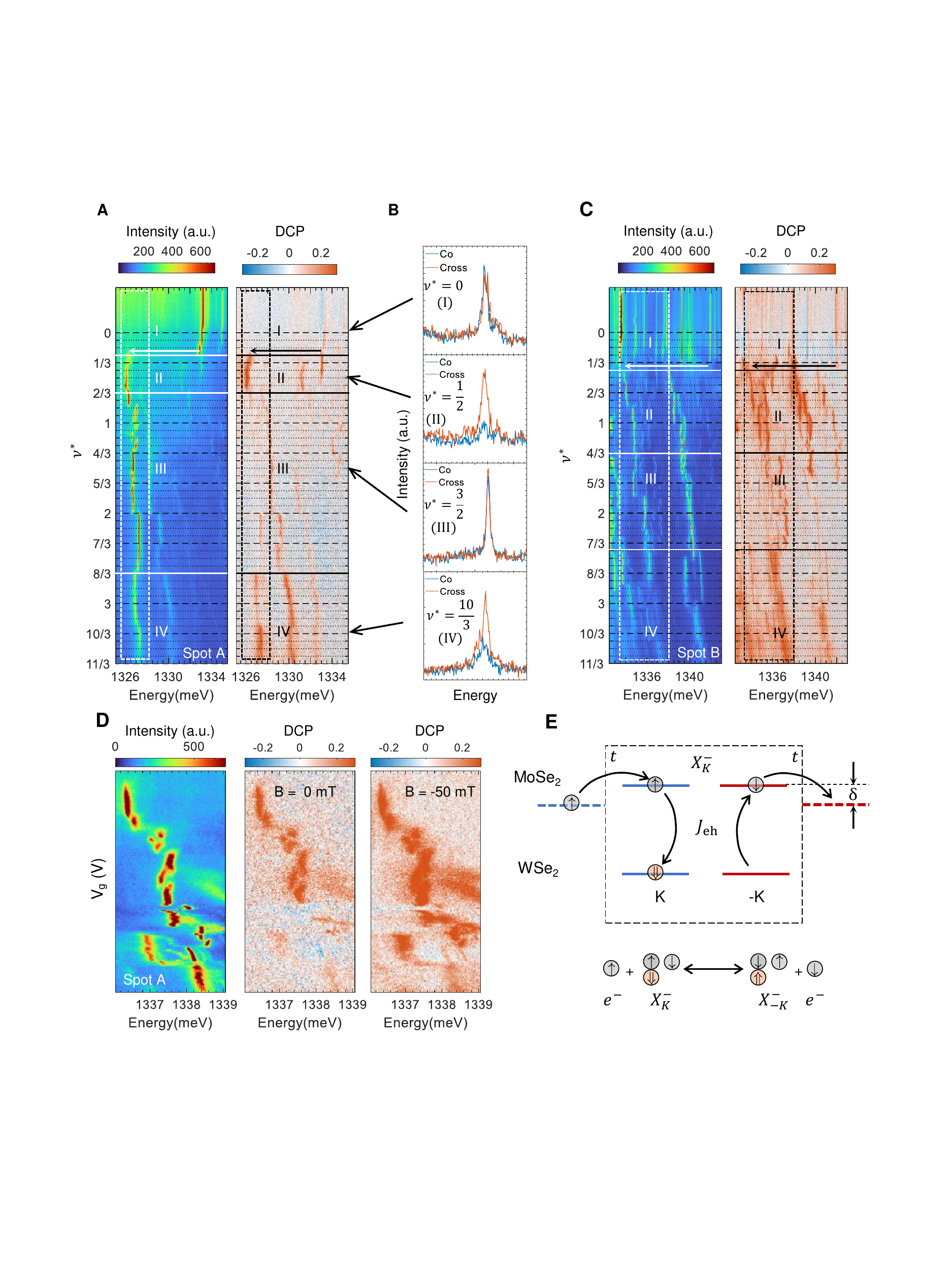}
\end{figure}
\clearpage

\noindent {\bf Fig. 4. Doping dependent valley polarization of moir\'e excitons.} {\bf(A)}
PL intensity (left) and degree of PL circular polarization DCP (right) as a function of fractional fillings at spot A. The filling for the DCP data is separated by white dashed lines into four regions: (I) unpolarized intrinsic region; (II) cross-polarized trion region red-shifted by 7 meV from intrinsic counterparts; (III) unpolarized doped region; (IV) cross-polarized doped region with broad linewidth and weak intensity. The arrows indicate the 7 meV redshift and the boxes outline excitons that we focus on. {\bf(B)} Helicity-resolved PL spectra at four fillings, $\nu^*$ = 0, 1/2, 3/2, 10/3. {\bf(C)} PL intensity (left) and DCP (right) change with fractional fillings at spot B, showing similar behavior as spot A. {\bf(D)} The PL intensity (left) and DCP at 0 T (middle) and -50 mT (right). The small magnetic field recovers the DCP, which is a signature of e-h exchange interaction quenching. {\bf(E)} Schematic of resonant tunneling induced exchange process of a K-valley moir\'e trion with a nearby singly-occupied moir\'e site at intermediate doping. The valley-conserving resonant tunnelings of moir\'e electron and trion electrons, denoted by tunneling amplitude $t$, together with e-h exchange $J_\mathrm{eh}$ enable an effective exchange interaction which reduces the DCP. The amplitude of this process, which is second-order in $t$, is enhanced at resonance when $\delta \approx$ 0. The net effect of this process is a spin flip-flop process between the moir\'e electron and the trion hole. The (hermitian) conjugate of this process is not depicted here for brevity (see Supplementary text).



\section*{Supplementary material}

## Methods

### Device fabrications

The dual-gated transition metal dichalcogenide heterotrilayer devices were fabricated via layer-by-layer dry transfer method using a polycarbonate (PC) stamp [32]. The $WSe_2$ and $MoSe_2$ monolayers, few-layer graphene and thick hBN are first mechanically exfoliated from bulk crystals on 300 nm $SiO_2$/Si substrates. The thickness of flakes was then determined by their optical contrast, from which we estimate the hBN thickness to be around 200 nm, consistent with atomic force microscopy measurements. The reason to select thick hBN is to reduce the effect of charge fluctuations near the sample. For the sample stacking process, we use a PC stamp to pick up (in sequence): the few-layer graphene top gate, top hBN flake, top $WSe_2$ monolayer, $MoSe_2$ monolayer, bottom $WSe_2$ monolayer, the few-layer-graphene contact, the bottom hBN flake and the few-layer graphene back gate. To produce small twist angles between the top $WSe_2$ and the middle $MoSe_2$ layers, as well as the bottom $WSe_2$ and middle $MoSe_2$, we use the tear-and-stack method [33] to pick up half of a $WSe_2$ flake, align with the $MoSe_2$ layer within 1° uncertainty and pick up the other half of the $WSe_2$ flake with a rotation angle of 3° (Device 1, main text and Fig. S1-S20) and 60° (Device 2, Fig. S21) relative to the top $WSe_2$. The complete, encapsulated heterostructure is finally released onto a 300 nm $SiO_2$/Si substrate with pre-patterned electrodes (5 nm Cr/55 nm Au) at 170 °C.

### Optical measurements and electrostatic gating

Photoluminescence and reflection contrast spectroscopy measurements were performed with Device 1 inside a closed-cycle optical cryostat (BlueFors cryogenics, 4 K) equipped with a ± 8 T magnet in Faraday geometry. Device 2 was measured inside another cryostat (AttoDry800, 7 K). A piezoelectric controller (Attocube systems) is used to position the sample. The excitation laser for the photoluminescence measurements is a mode-hop-free tunable continuous-wave Ti:Sapphire laser (MSquared Lasers) with a wavelength resolution of 0.1 pm, while a halogen lamp serves as the white light source for the reflectance contrast measurements (Thorlabs SLS201L). The laser/white light beams are focused on the sample with a spot size of ~1 μm using an achromatic objective (NA = 0.42 for AttoDry 800 and NA = 0.63 for BlueFors cryogenics). The photoluminescence emission or reflected white light is collected through the same objective and directed to a high-resolution (focal length, 500 mm for AttoDry 800 and 750 mm for BlueFors cryogenics) spectrometer (Princeton Instrument HR-500 for AttoDry 800 and Princeton Instruments SP-2750i for BlueFors cryogenics) where it is dispersed by a 1,200 grooves per mm or 300 grooves per mm grating (both blazed at 750 nm). A charge coupled device (Princeton Instrument PIXIS-400 CCD for AttoDry 800 and PyLoN CCD for BlueFors cryogenics) is used as a detector. The polarization of the incident laser is controlled by a polarizer and a liquid crystal variable retarder. The polarization-resolved measurements are performed by using a $\lambda/4$ waveplate (achromatic, 690–1,200 nm) placed before a Wollaston prism. The $\sigma^+$ and $\sigma^-$ components of circularly polarized light are converted into s and p components of linearly polarized light and displaced by the Wollaston prism.

To tune the charge density or apply an electric field to the sample, voltages are applied to the graphene top gate and graphene bottom gate through Keithley 2400 source meters while the

graphene contact is grounded. The top and bottom gates are nearly symmetric with 180-200 nm hBN gate dielectrics, so negligible electric field was measured with symmetric gating voltages.

**DFT calculations**

The ab initio calculations were performed within the Vienna Ab initio Simulation Package (VASP) [34] using a projector-augmented wave (PAW) pseudopotential in conjunction with the Perdew–Burke–Ernzerhof (PBE) [35] functionals and plane-wave basis sets. For the trilayer heterostructures with $MoSe_2$ and $WSe_2$, we used a supercell containing 2979 atoms. The twisted angles for the top and bottom two layers are 3.15° and 0°, respectively, giving rise to a moiré length of 6.04 nm. Due to the huge supercell in our simulations, the Brillouin zone was sampled by a 1×1×1 Monkhorst–Pack k-mesh. A vacuum region of 15 Å was applied to avoid artificial interaction between the periodic images along the perpendicular direction. Because of the absence of strong chemical bonding between layers, van der Waals density functional in the opt88 form [36] was employed for structural optimization. All structures were fully relaxed until the force on each atom was less than 0.05 eV $Å^{-1}$.

As the moiré system is filled from the electron side, and there is a large band offset for 400 meV between monolayer $MoSe_2$ and $WSe_2$, we only consider the electron moiré bands from $MoSe_2$ layer with corrugation, which is extracted from the fully relaxed trilayer heterostructures. Spin-orbital couplings are included in the electronic calculations. We extract the orbital localizations of the DFT wavefunctions with the VASPKIT code [37].

# Supplementary Information

**Determination of moiré length $a_M$**

The moiré length is determined by the twist angle and lattice mismatch. Due to the small lattice mismatch (0.1%) of $WSe_2$ and $MoSe_2$, the moiré length mainly depends on the twist angle $\theta$ by $a_M \approx a_0/\theta$. However, for the trilayer structure with different top and bottom twist angles, the moiré potential has a beating pattern as shown in Fig. 1C or Fig. S16, S18, with various length scales between potential minima. To achieve an effective length scale, we first use the capacitor model to independently obtain the electron density. Then we assign $v^* = v = 1$, i.e., the voltage where we see the first kink in the global reflectance as half-filling of the first minima, to obtain a moiré density and thereafter the effective moiré length.

Based on the electric field induced energy shift $\Delta E$ of 3.5 meV by $V_{bg}$ change from 6 to 10 V (Fig. S5), the total hBN thickness $D_{hBN}$ is given by $\Delta E = e \cdot d \cdot \epsilon_g/\epsilon_w \cdot \Delta V_{bg}/D_{hBN}$, where the $\epsilon_w = 7.4$ is the dielectric constant of MoSe2/WSe2, $\epsilon_g = 3.7$ is the hBN out-of-plane dielectric constant, and $d = 0.7$ nm is the interlayer distance. The total hBN thickness $D_{hBN}$ is calculated to be 400 nm so that each hBN thickness is about 200 nm, consistent with the optical contrast estimation and AFM measurements. With the hBN thickness, we can estimate the electron doping density by $n = 2\epsilon_g\epsilon_0/eD_{hBN} \cdot 2\Delta V_g$. In Fig. 2&3, $\Delta v^*=1$ is corresponding to $\Delta V_g= 8.5$ V, so that the moiré density $n_0$ is equal to $1.74\times10^{12}$ cm$^{-2}$. The effective moiré periodicity is calculated to be $\overline{a_M} = 1/\sqrt{n_0 \sin\left(\frac{\pi}{3}\right)} \sim 8$ nm.

Another independent method to estimate the mean moiré length is the transformation of the energy shifts of dipolar excitons to be a length scale $r_{eff}$. The energy shifts of a dipolar exciton by one electron at $r_{eff}$ (Fig. S15A) is $\Delta E = 1/4\pi\epsilon_r\epsilon_0 \cdot (1/r_{eff} - 1/\sqrt{r_{eff}^2 + d^2})$, which is a blueshift of the exciton energy. The $\epsilon_r = 4.95$ is the dielectric constant of hBN with both in-plane and out-of-plane contributions. The explanation of the redshifts in the experimental data is that electrons occupied the site near the dipolar exciton becomes unoccupied due to reconstruction of the electron crystals (Fig. S15B). Considering the spectral jumps in the data caused by one electron moving in and out of the dipolar exciton sensing range, which is 14 nm for 26 µeV spectral resolution, the $r_{eff}$ distribution from $\Delta E$ distribution is shown in Fig. 3G. The average of all $r_{eff}$ is ~7.4 nm, consistent with the above capacitor model estimation.

**Correction for the doping dependence after subjecting to high voltage.**

The gate dependence range of the device 1 is permanently changed after applying $V_g = 40$ V to the sample, however, the spectral jump dynamics remain the same as shown in Fig. S8. Mapping the voltage before the 40 V gate scan (Fig. S8A) to the voltage after the 40 V gate scan (Fig. S8B) for each energy shift, the replotted gate scan after 40 V scan (Fig. S8C) is almost the same as before. Besides, it is worthy to be noted that Fig. 2A and 3B are different data sets before the 40 V scan with different voltage scan range (10 V for Fig.2A and 20 V for Fig.3B) but show the same energy shifts at similar voltages. Fig. S6B, C, D have different scan ranges after the 40 V scan but also show the same energy shifts at similar voltages. Therefore, it suggests 40 V changes the gate instead of sample properties, in other words, electronic crystal states are unchanged. For the fraction analysis, we use the data before 40 V scan. The filling fractions in

Fig. S11, S12, S13 are assigned according to one-to-one mapping between voltage sets before and after applying 40 V.

**Monte Carlo simulations of long-range charge ordered states on multi-orbital lattices.**

To simulate the electronic crystals at various fractional fillings, we employ a classical simulated annealing scheme based on minimizing the electrostatic energy of electrons that we employ for the classical charges on multi-orbital lattices. We use a screened Yukawa-like potential in the form of $V(r_{ij}) = \frac{e^2}{4\pi\epsilon_r\epsilon_0 r_{ij}} e^{-r_{ij}/r_0}$. Here $r_0 = N/2$ is the interaction length scale, where N is the length of the supercell that we consider. For the $3°/0°$ trilayer, the sites for electrons are the moiré potential minima determined from the DFT calculations, consisting of MX as the first orbital (#1, -30.4 meV, spin down) and MM as the second orbital (#2, -3.4 meV, spin down) (Fig. S16). For the $1°/4°$ trilayer (Fig. 1C) or $2°/3°$ trilayer (Fig. S18), the electron moiré potential is calculated by $V_e = -(D^t(\mathbf{R}) + D^b(\mathbf{R}))/2$. Here $D^{t/b}(\mathbf{R})$ is the top/bottom exciton potential, given by $D^{t/b}(\mathbf{R}) = D_0 f_0^{t/b}(\mathbf{R}) + D_{+1} f_{+1}^{t/b}(\mathbf{R}) + D_{-1} f_{-1}^{t/b}(\mathbf{R})$, where $f_m^{t/b}(\mathbf{R}) = 1/9 \left| e^{-i\mathbf{K}^{t/b}\cdot\mathbf{R}} + e^{-i(\hat{C}_3\mathbf{K}^{t/b}\cdot\mathbf{R} - m\frac{2\pi}{3})} + e^{-i(\hat{C}_3^2\mathbf{K}^{t/b}\cdot\mathbf{R} + m\frac{2\pi}{3})} \right|^2$ with $\mathbf{K}^{t/b}$ being the wavevector at the corner of top and bottom Brillouin zones. The three parameters $(D_0, D_{+1}, D_{-1})$ = (20, -8, 83) meV for WSe$_2$/MoSe$_2$ heterostructure[19]. The calculated electron moiré potential shows the degenerate local minima for each unit cell, which is another signature of multi-orbital lattices. To avoid the boundary effect, we use $9 \times 9$ replicas for the electron total energy calculations. The simulated electron configurations are shown in Fig. S15, S17D, S18B, supporting that the electrons with interactions form crystal-like structures. For the dipolar exciton spectral jumps, we consider the excitons at the MX position in $3°/0°$ trilayer and bottom exciton potential minima in $1°/4°$ or $2°/3°$ trilayers. The simulated exciton energy is achieved by summing the electron-dipole energy over all filled electrons, capturing the experimental features of red- and blue-shifts with overall blueshift (Fig. S17A, S18C). One feature of the multi-orbital is also captured in the simulation is the jump size variation with the orbital, i.e., the width of the jump size distribution is different, as the length scale for different orbital is different (Fig. S17A, C). The experimental histogram in Fig. S17B shows the signature of the multi-orbital by the asymmetry between $\nu^* > 3/2$ and $\nu^* < 3/2$.

**Exchange Hamiltonian for moiré trions**

The valley mixing process for moiré trions includes the valley flip of exciton by the electron-hole exchange interaction $J_{eh}$, and valley-conserving resonant tunneling $t$ of both an electron occupying a nearby moiré site and the trion electron with detuning $\delta$. The Hamiltonian of the process is

$$H = J_{eh} t^2/\delta^2 \left( \hat{X}_{-K}^+ \hat{c}_{-K}^+ \hat{c}_K \hat{X}_K + h.c. \right)$$

where $\hat{X}_K = \hat{e}_{-K} \hat{e}_K \hat{h}_K$ is the annihilation operator for the K-valley trion with the annihilation of K valley exciton $\hat{e}_K \hat{h}_K$ and -K-valley trion electron $\hat{e}_{-K}$. $\hat{c}_K$ is the annihilation operator for the electron in the moiré site. The amplitude of the process is $J_{eh} t^2/\delta^2$, which depends on both $t$ and $\delta$. If the detuning $\delta$ is close to zero, the effect is enhanced to reduce the DCP of trion. The net effect of this process is a spin flip-flop process between the moiré electron and the trion hole.

4

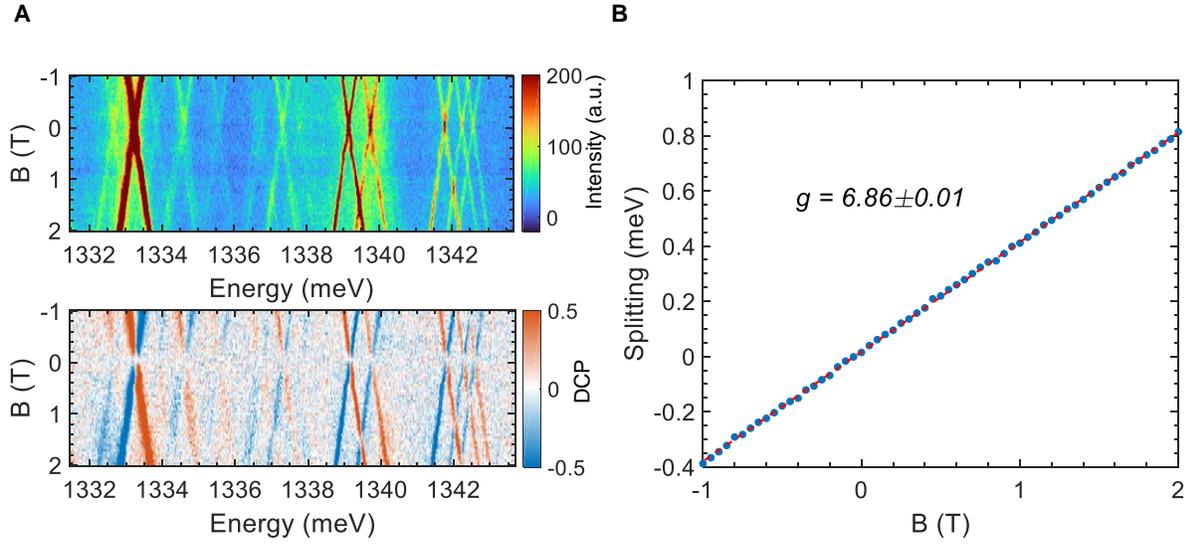

**Fig. S1. Magneto-PL dependence of bottom dipoles.** (A) The PL intensity (top panel) and degree of circular polarization (DCP) (bottom panel) as the magnetic field $B$ is swept from -1 to 2 T. (B) Zeeman splitting of an emitter versus $B$ yields a g-factor of 6.86. The linear Zeeman splitting has no observable fine structure and a g-factor of 6.86 is consistent with moiré exciton features in $WSe_2/MoSe_2$ close to 0º twist angle [23]. The cross-polarized feature under tiny $B$ field in Fig. 1G also confirms the assignment. The excitation laser is linearly polarized with the energy $E = 1.70$ eV and power $P = 100$ nW.

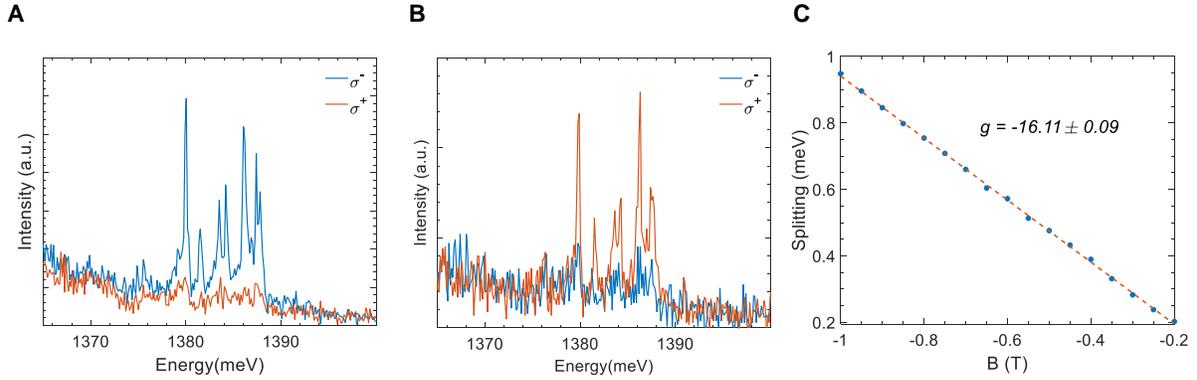

**Fig. S2. Helicity-resolved PL spectra for the neutral top dipoles.** PL spectra under (A) $\sigma^-$ and (B) $\sigma^+$ excitation show that the neutral top dipoles are co-polarized at $\boldsymbol{B}$ = 0. (C) The Zeeman splitting versus $\boldsymbol{B}$ under linear excitation gives a g-factor of -16.11. The co-polarization without magnetic field and g-factor of -16.11 suggest that the top $WSe_2/MoSe_2$ has close to 60º twist angle[*23*]. This is possibly because the top $WSe_2$ is unintentionally folded at the edge. The excitation laser is at 1.70 eV and 600 nW.

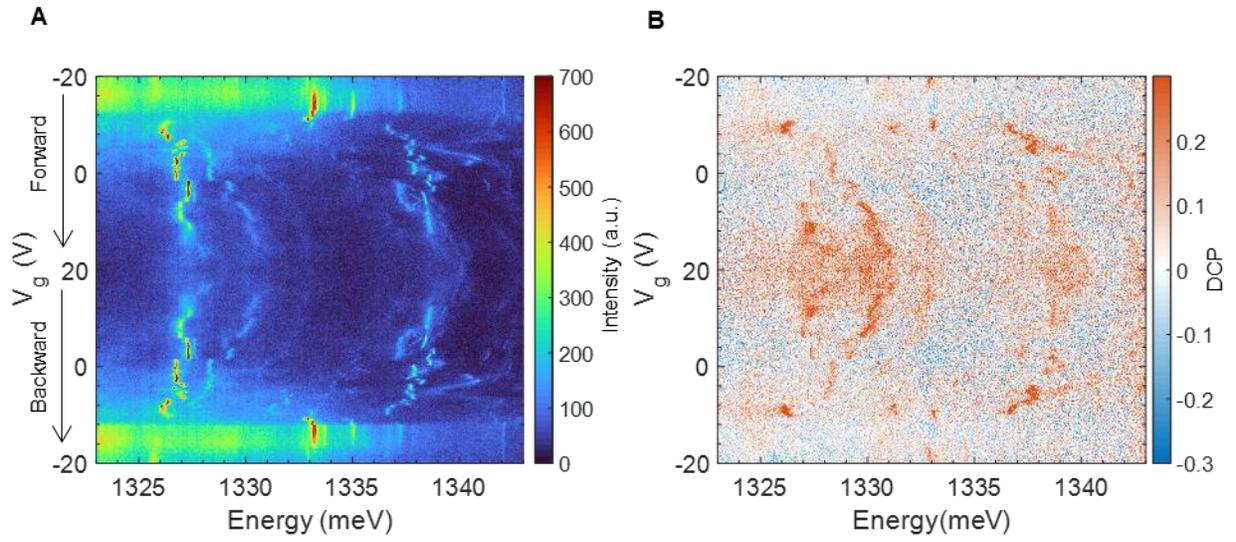

**Fig. S3. Reproducible energy shifts with electron doping at spot A.** (A) PL intensity and (B) DCP of localized interlayer excitons change with $V_g$. When the applied gate voltage is reversed, the energy shifts and corresponding DCP are reversed. The excitation laser is $\sigma^-$ polarized at 1.70 eV and 50 nW.

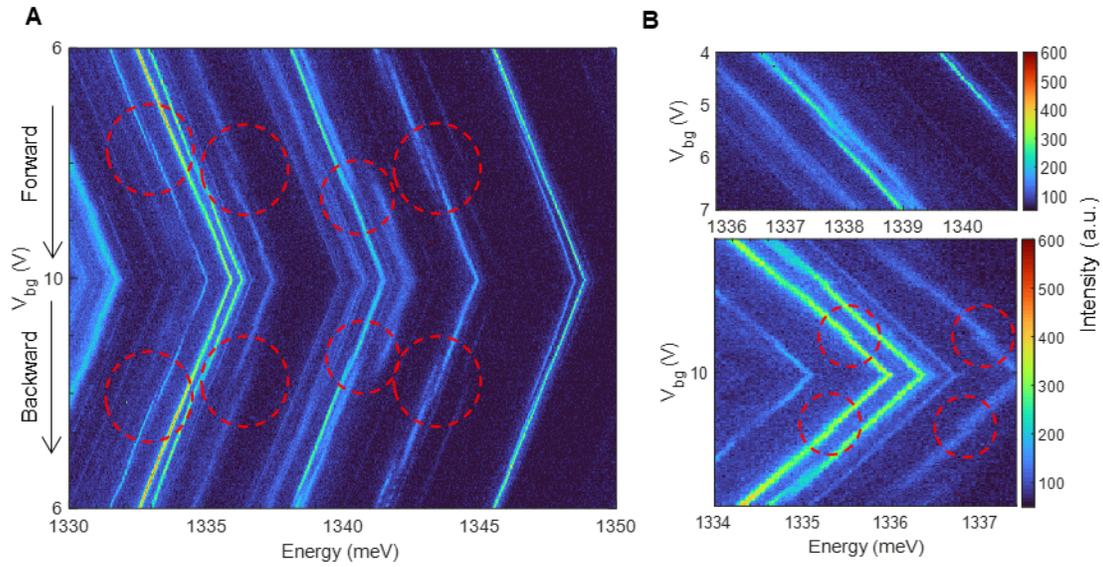

**Fig. S4. Reproducible energy shifts with doping and electric field at spot B.** (A). Back gate dependence of PL intensity forward from 6 V to 10 V and backward from 10 V to 6 V. When the gate voltage is reversed, the energy shifts are reversed. (B). Zoomed in examples of the energy shifts. The excitation laser is at 1.70 eV and 50 nW.

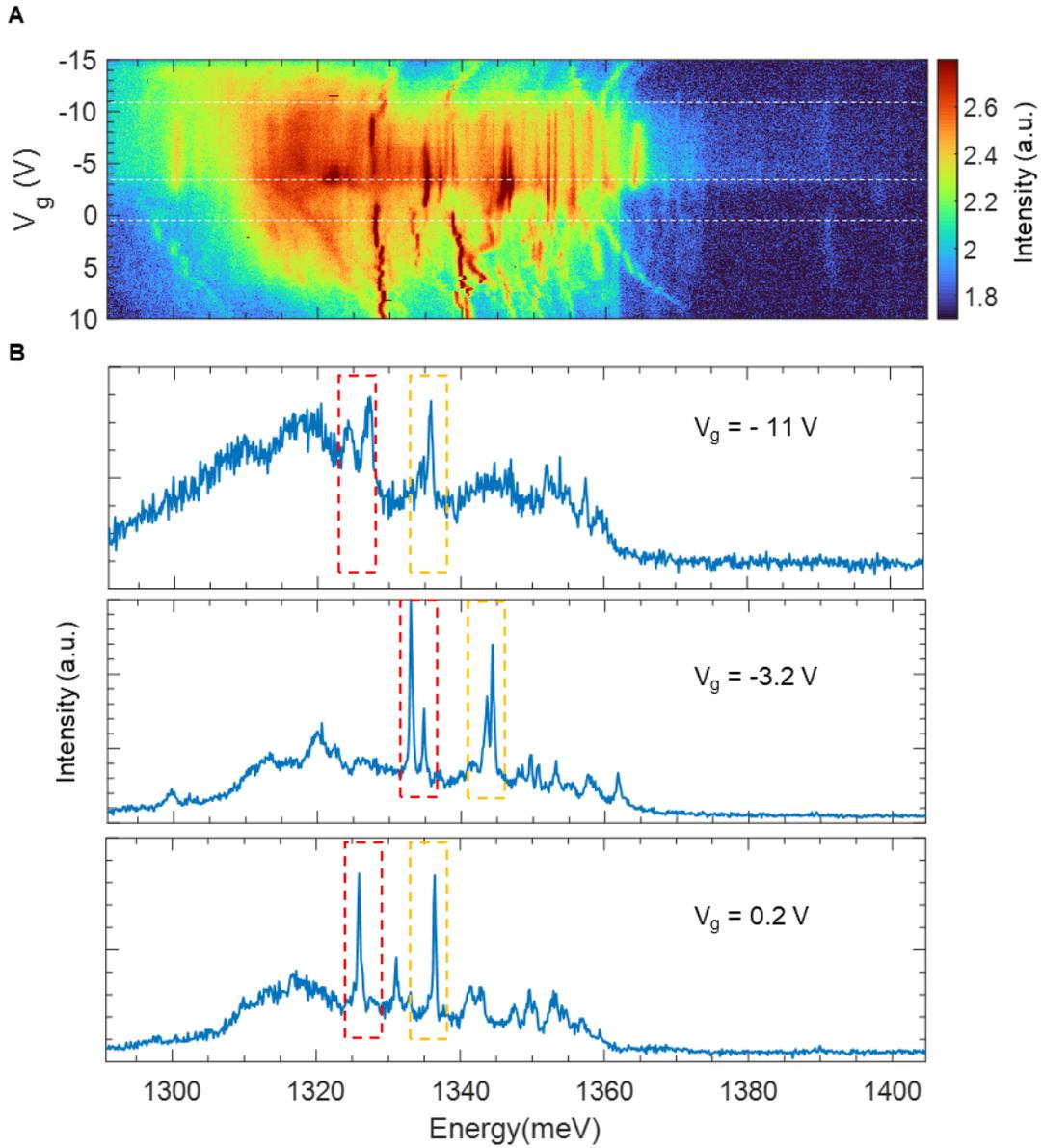

**Fig. S5. Formation of moiré trion with electron and hole doping.** Gate dependence of PL emission of localized interlayer excitons (A) and lineplots at certain voltages (B) show 6-7 meV redshifts of exciton energy from intrinsic (-3.2 V) to hole doping (-11 V) and electron doping (0.2 V). The red and yellow dashed squares are the examples of moiré excitons.

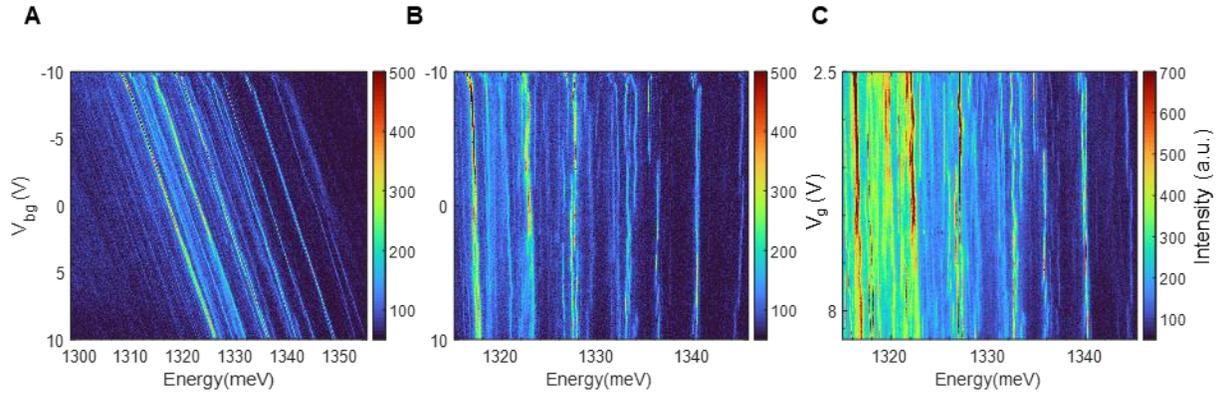

**Fig. S6. Comparison between the gate dependences at spot B.** (A). Gate dependence by applying the voltage to the back gate together with other layers grounded, giving rise to both electric field and doping effect to the sample. (B). Reconstructed gate dependence by removing the slope in (A). (C). Gate dependence by applying the same voltages to the back gate and top gate. The excitation laser is at 1.70 eV and 50 nW.

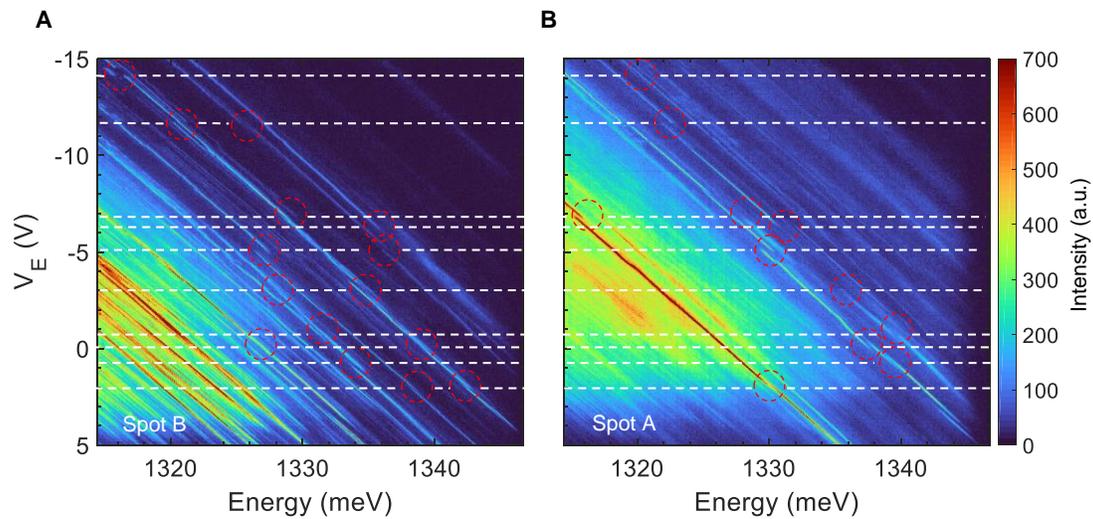

**Fig. S7. Correlated energy shifts between spot A and spot B with both electric field and doping effect.** The white dashed lines indicate the voltages where the energy shifts (red dashed circles) occur for both spot A and B. The excitation laser energy is 1.70 eV and power is 80 nW.

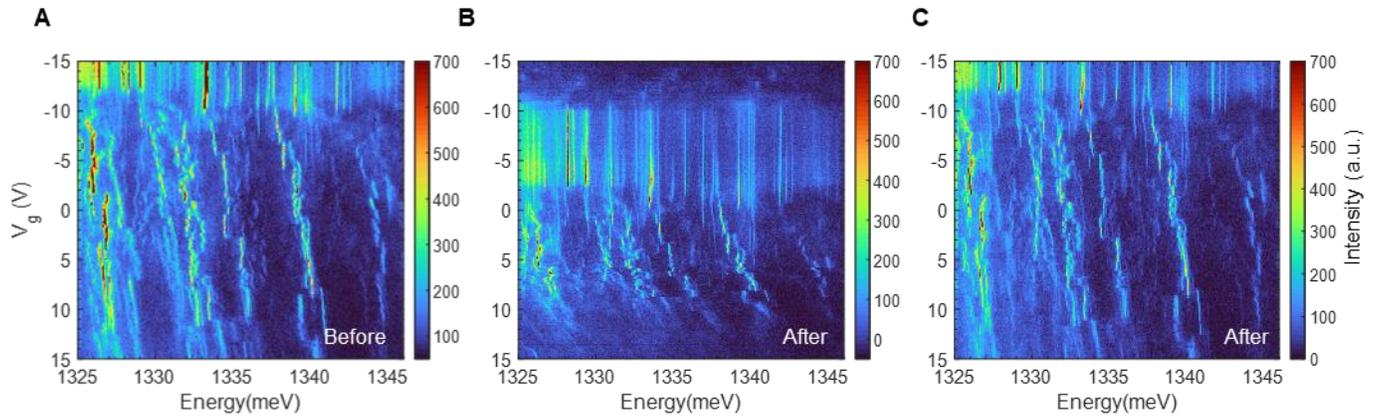

**Fig. S8. Correction for the gate dependence after high voltage scan.** (A). Gate dependence at spot B before applying 40 V to the sample. (B). Gate dependence at spot B after applying 40 V to the sample. The charge neutral point is shifted and voltage range for electron doping is smaller compared to before. (C). Reconstructed gate dependence from (B) via one-to-one mapping between voltages for energy shifts in (A) and (B). For further details, see supplementary text. The excitation laser is at 1.70 eV and 50 nW.

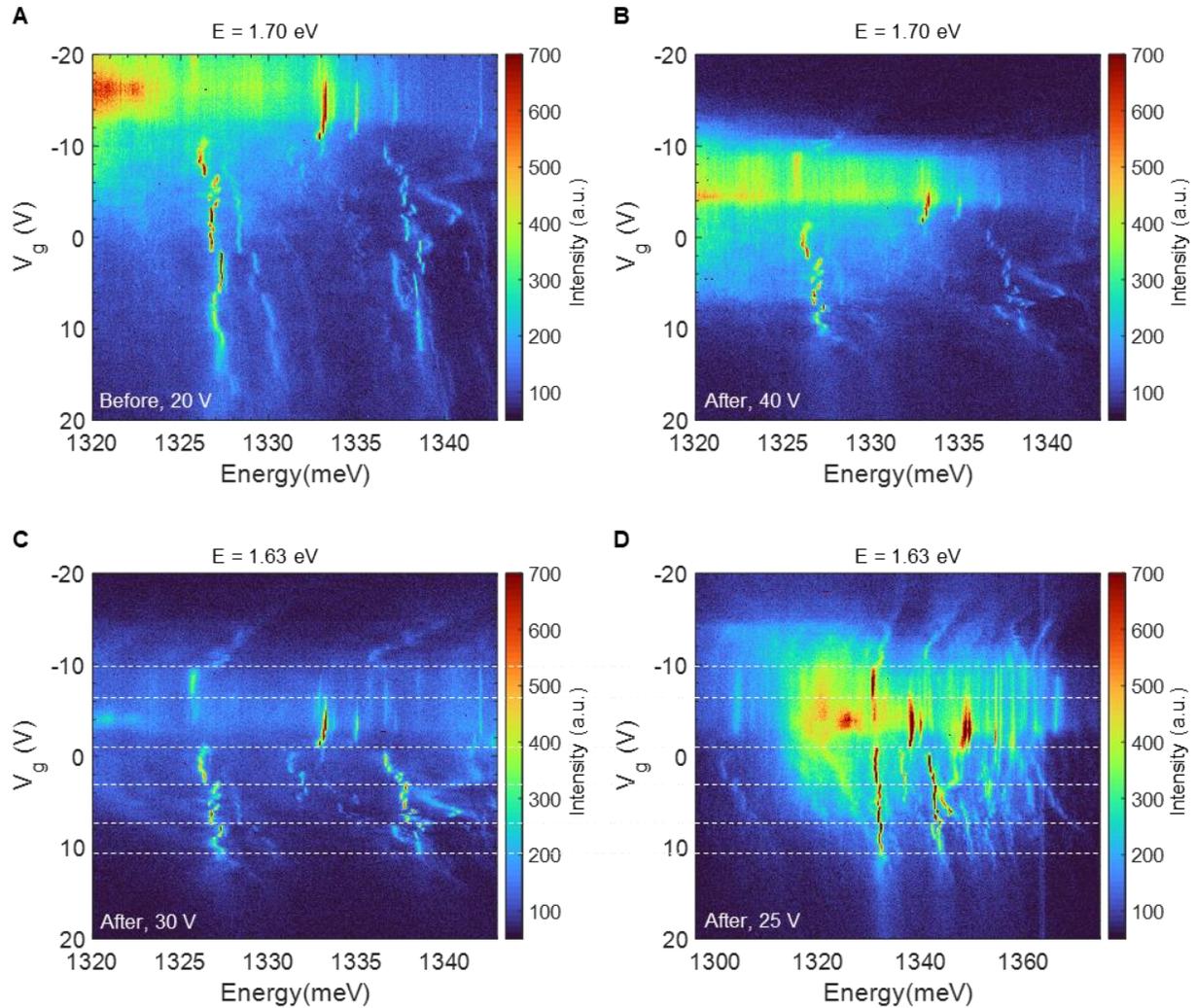

**Fig. S9. Comparison between the gate dependences at spot A.** (A). Gate dependence before applying 40 V to the sample. (B-D). Gate dependence after applying 40 V to the sample. The voltage scan range is +/- 20 V (A), +/- 40 V (B), +/- 30 V (C), +/- 25 V (D), respectively. The similar energy shifts occur in (C) and (D) at the white dashed lines. The excitation laser is at 1.70 eV (WSe$_2$ resonance) and 70 nW for (A-B) while it is at 1.63 eV (MoSe$_2$ resonance) and 100 nW for (C-D). The spectral resolution is 1200 grooves/mm for (A-C) and 300 grooves/mm for (D).

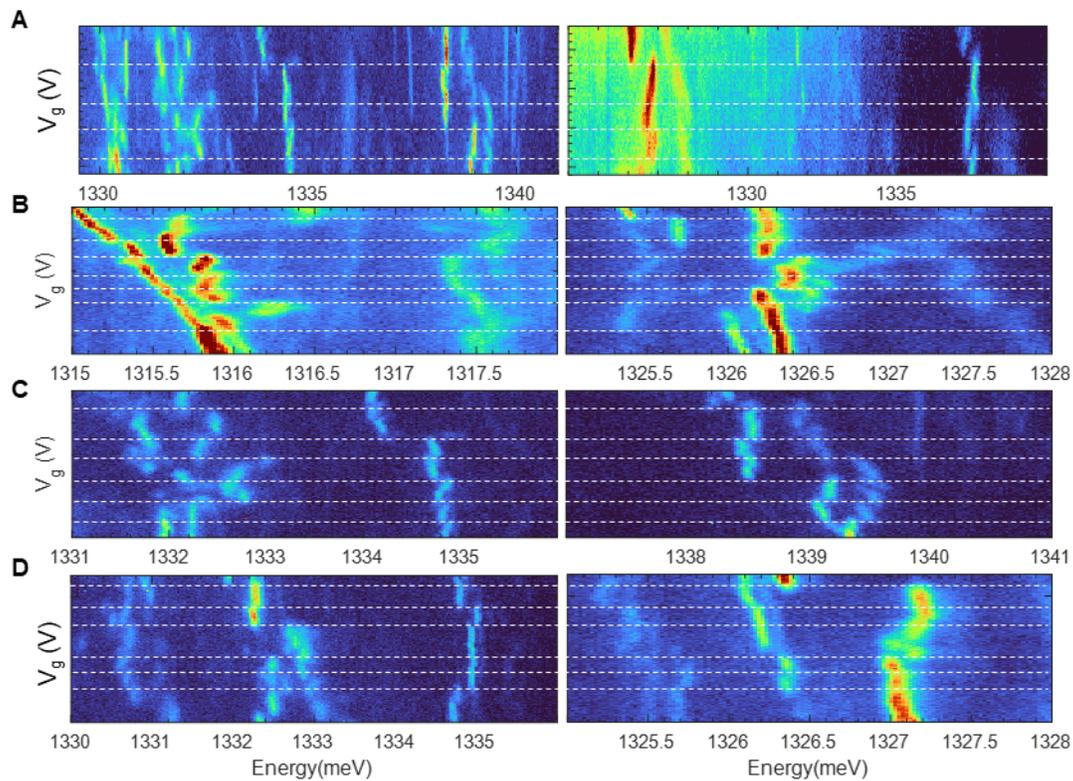

**Fig. S10. Doping examples.** (A). The correlated spectral jumps between spot A (right) and spot B (left) landing on the white lines. (B-D). The correlated spectral jumps within spot B landing on the white lines.

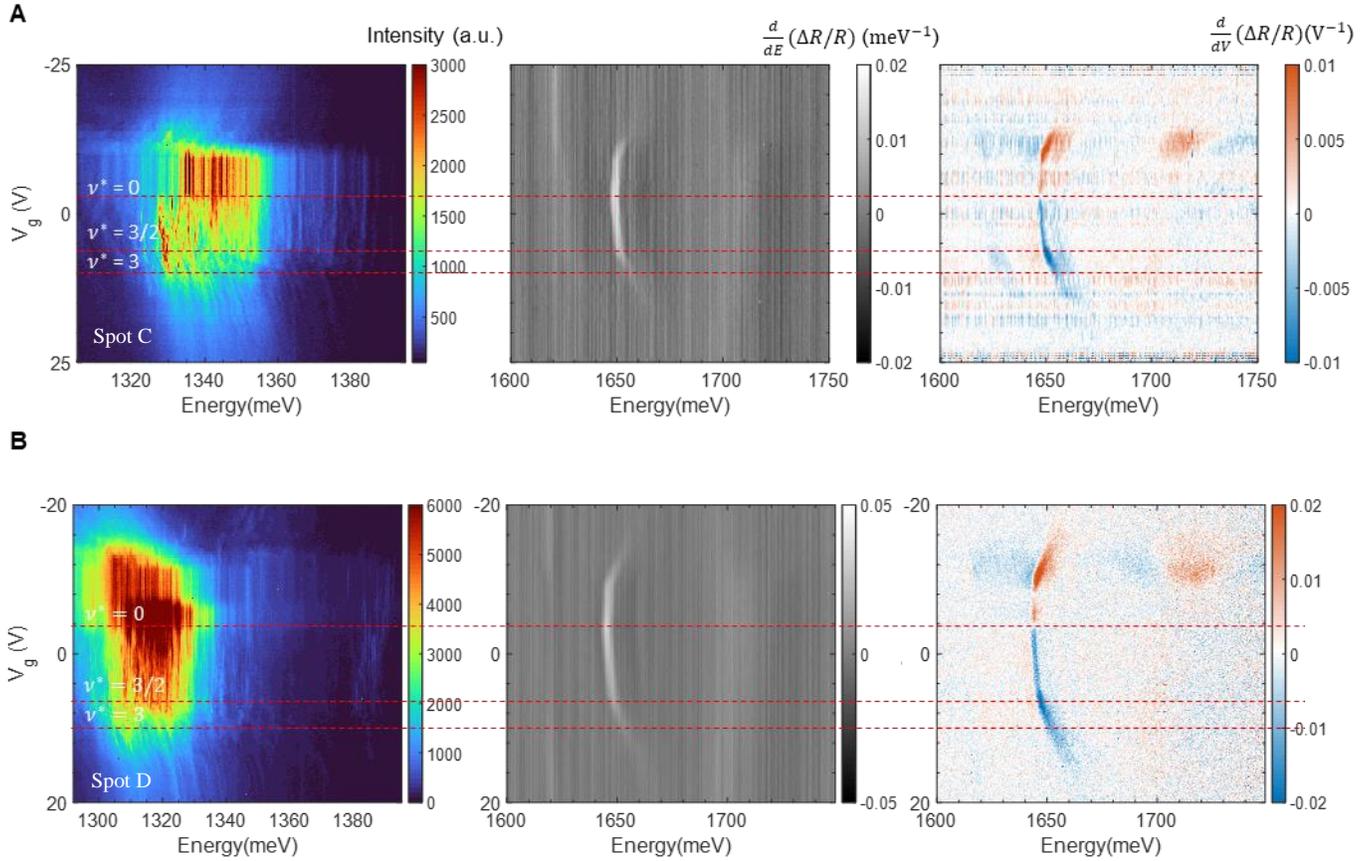

**Fig. S11. Correlation between PL and reflectance contrast spectra at spot C and spot D.**
(A). Gate dependence of PL (left panel), derivative of reflectance contrast over energy (middle panel) and derivative of reflectance contrast over voltage (right panel) at spot C. (B). Gate dependence at spot D. From intrinsic region $\nu^* = 0$ to electron doped region $\nu^* = 3/2$, localized interlayer exciton energy shifts with increased electron density and their intensity decreases a lot at $\nu^* = 3$, which is consistent with the MoSe$_2$ resonance blue-shifting, broadening and almost disappearing at $\nu^* = 3$ in the reflectance contrast spectra. The excitation laser energy is 1.70 eV and power is 200 nW for (A) and 600 nW for (B).

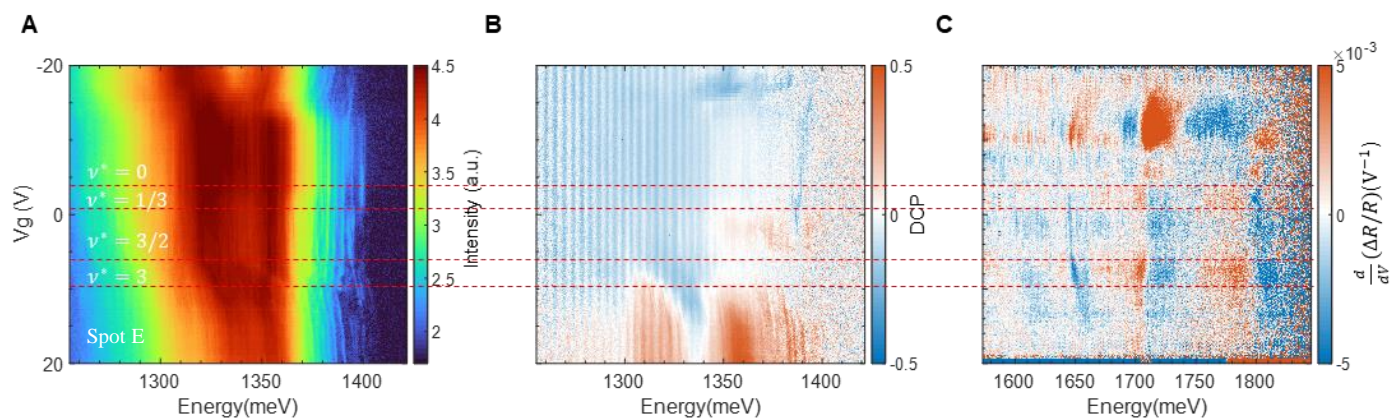

**Fig. S12. Correlation between PL and reflectance contrast spectra at spot E.** Gate dependence of PL (A) and DCP (B) of interlayer excitons, which have broad linewidth. (C). Gate dependence of reflectance contrast. At $\nu^* = 1/3$, the DCP suddenly increases. From $\nu^* = 1/3$ to $3/2$, the DCP stays constant for a certain range and then decreases to zero. From $\nu^* = 3/2$ to 3, the DCP starts to recover. This process correlates with features in the reflectance contrast spectrum. The excitation laser is $\sigma^-$ polarized at 1.70 eV and 5 $\mu$W.

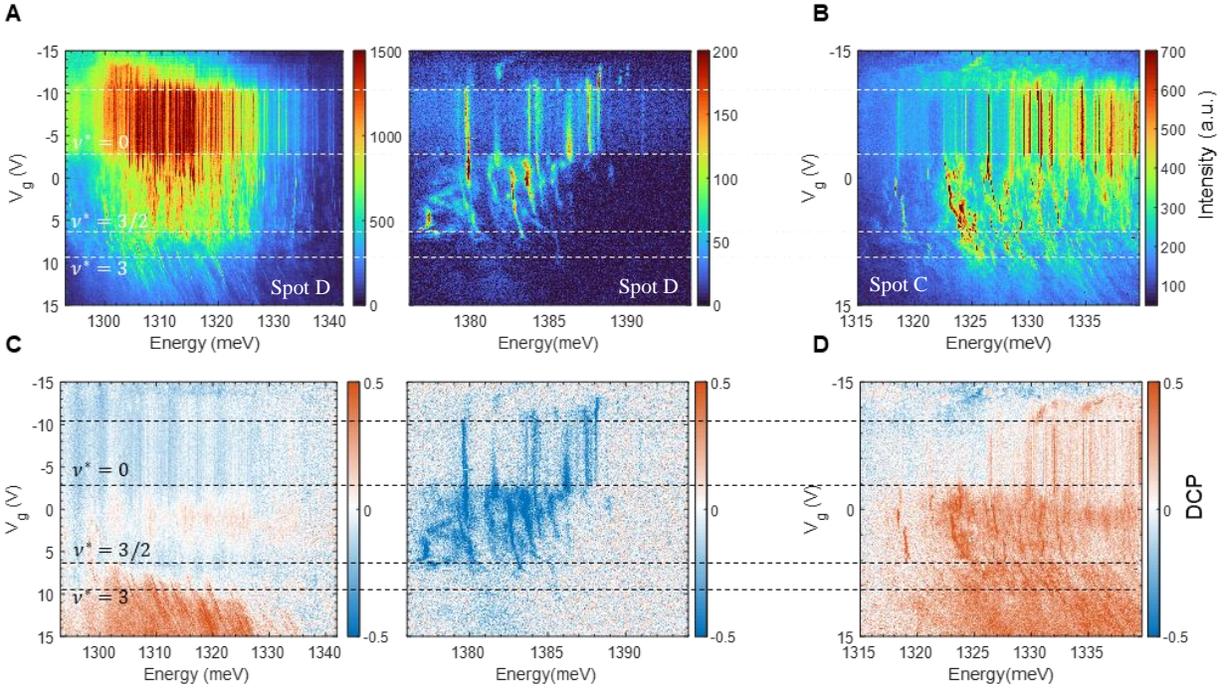

**Fig. S13. Correlation of PL spectra between spot C and spot D.** Gate dependence of PL (A) and DCP (C) for bottom dipoles (left panel) and top dipoles (right panel) at spot D correlates to that of PL (D) and DCP (D) at spot C. At $\nu^* = 0$, the PL spectra change from not shifting in the intrinsic region to exhibiting sudden red and blue shifts in the electron doped region for both dipoles and sample positions. At $\nu^* = 3/2$, top dipoles suddenly become weak, and the DCP of bottom dipoles starts to recover at spot D. In addition, the sudden energy shifts for localized interlayer excitons happen for both positions at $\nu^* = 3/2$. The overall blueshifts of both top and bottom localized dipolar excitons exclude the possibility of electric field effect. The excitation laser is $\sigma^-$ polarized at 1.70 eV.

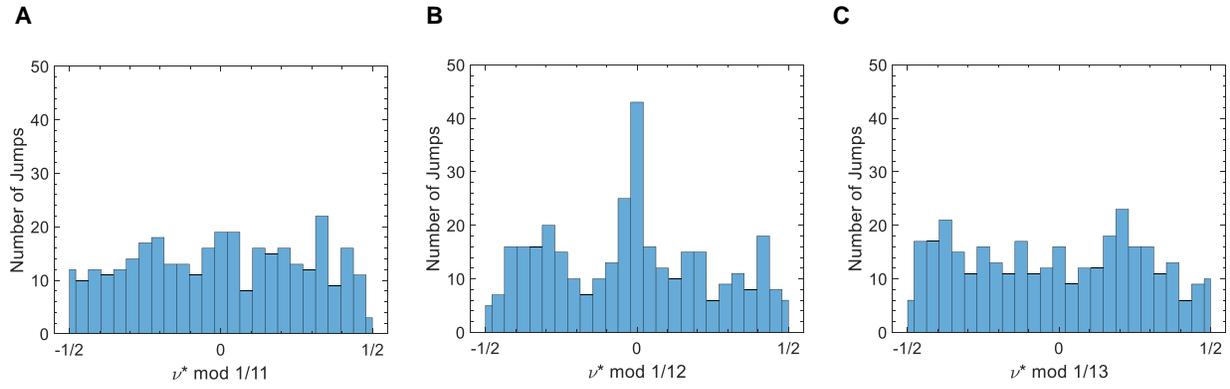

**Fig. S14. Histograms of residuals of fillings at which energy jumps occur with different moduli.** (A). Histogram of jump positions modulo 1/11. (B). Histogram of the jump positions modulo 1/12. (C). Histogram of the jump positions modulo 1/13. Only modulus 1/12 gives a peak at 0, which means fractions with denominator 12 are the main filling fractions.

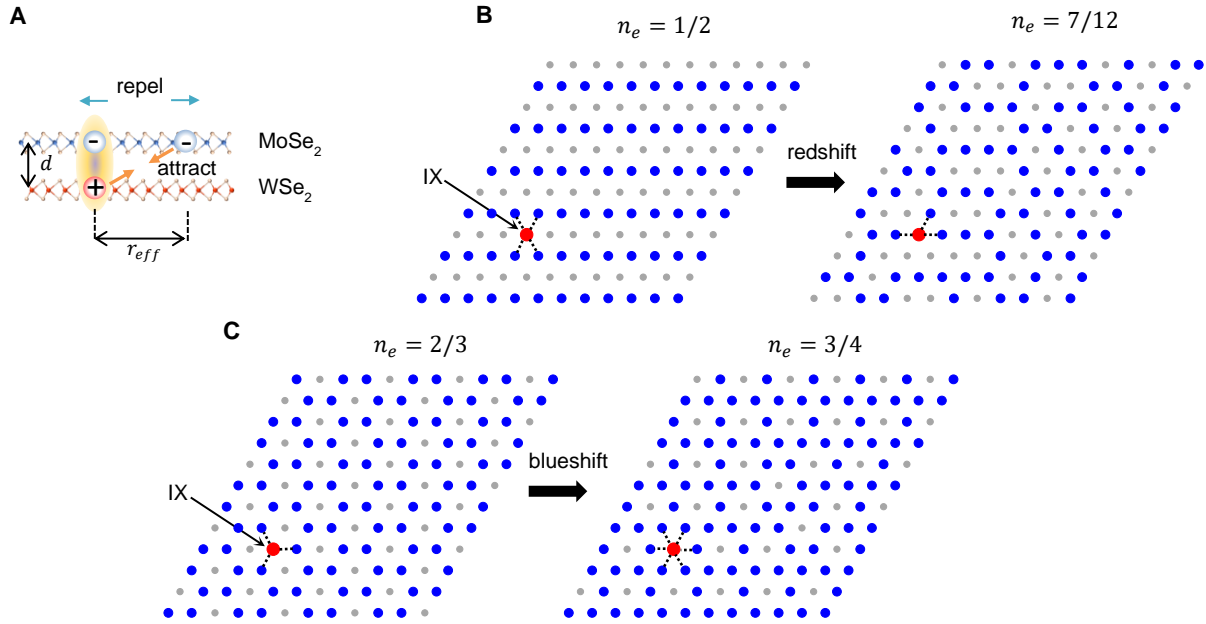

**Fig. S15. Examples of charge-ordered state transitions with red- or blue- shifts.** (A). Electrostatic model for the dipolar exciton energy change caused by the electron. With the repulsion between the electrons and attraction between the electron and hole, the form of exciton energy shift is given by $\left(1/r_{eff} - 1/\sqrt{r_{eff}^2 + d^2}\right)/4\pi\epsilon_r\epsilon_0$, which is a blueshift. (B). Dipolar exciton (IX) energy red shifts from filling $n_e = 1/2$ to $n_e = 7/12$. (C). Dipolar exciton energy blue shifts from filling $n_e = 2/3$ to $n_e = 3/4$. $n_e$ is the number of electrons per unit cell. Red dots are interlayer excitons, blue dots are occupied sites by electrons, and the grey dots are unoccupied sites. The black dotted lines indicate the nearest neighbor (NN) interactions.

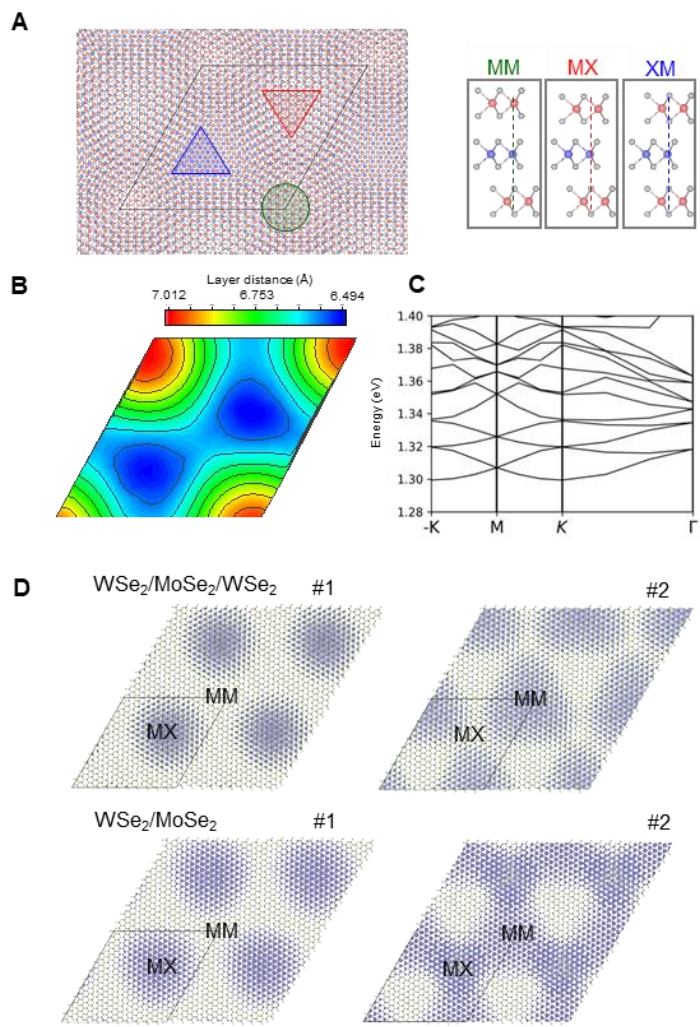

**Fig. S16. DFT calculations of band structure and electron localizations.** (A). Relaxed WSe$_2$/MoSe$_2$/WSe$_2$ trilayer structure with twist angles of 3.15° and 0° for the top and bottom bilayer, respectively. MM, MX, and XM are the three high-symmetry stacking locales. The unit cell length is 6.04 nm. (B). Corrugation of the trilayer heterostructure. (C). Bands of the trilayer with spin-orbital-coupling. (D). Electron localizations in the MoSe$_2$ layer for WSe$_2$/MoSe$_2$/WSe$_2$ (top panel) and WSe$_2$/MoSe$_2$ (bottom panel). For the trilayer, the first orbital (#1, -30.4 meV, spin down) and second orbital (#2, -3.4 meV, spin down) are localized at MX and MM, respectively; however, the second orbital of bilayer is localized at MX the same as the first orbital.

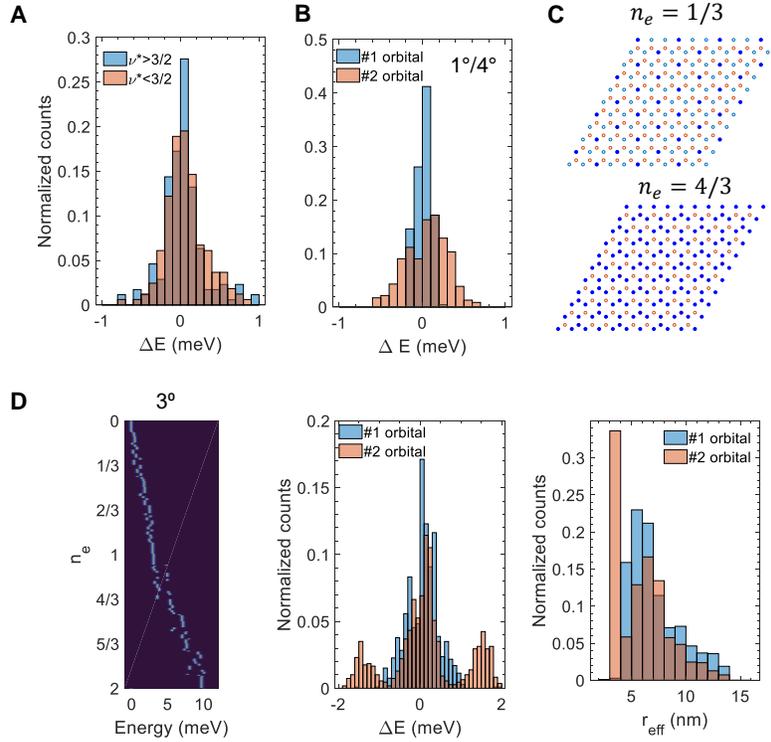

**Fig. S17. Monte Carlo simulation of charge-ordered states for electrons in 3°
heterostructure with two orbitals.** (A). The histogram of all the experimental energy shifts in
spot A and spot B, $\nu^* < 3/2$ has a smaller distribution than $\nu^* > 3/2$, suggesting a multi-orbital effect
as simulated in (B-D). (B). 1°/4° has different distributions for two orbitals. The second orbital
energy shifts are more broadly distributed than the first orbital. (C). $n_e = 1/3$ and $n_e = 4/3$ for the
3° heterostructure with two orbitals. Blue empty dots are unoccupied first orbital sites, red empty
dots are unoccupied second orbital sites, and blue solid dots are sites occupied by electrons.
When the first orbital is fully filled, electrons will start to fill the second orbital. (D). Simulation
of the dipolar exciton energy shifts with electron doping. The left panel shows red- and blue-
shifts with increased electron density, and an overall blueshift of 10 meV. $n_e = 0$ to 1 is the first
orbital and $n_e = 1$ to 2 is the second orbital. The middle panel is the histogram of energy shifts.
The second orbital energy shifts are more broadly distributed than the first orbital. The right
panel is the histogram of $r_{eff}$ extracted from energy shifts according to Fig. S15A. The second
orbital has a smaller $r_{eff}$ than the first one.

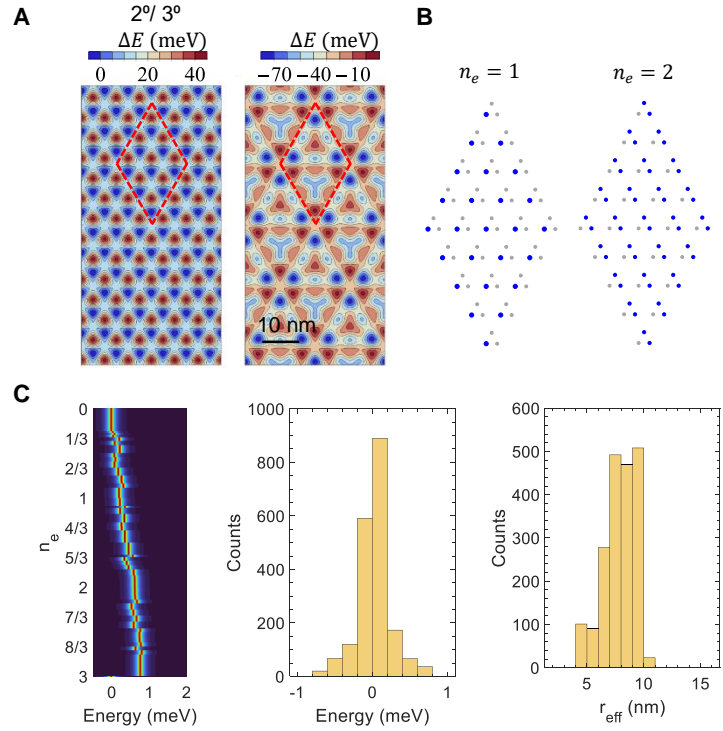

**Fig. S18. Monte Carlo simulation of charge-ordered states for electrons in the 2°/3° WSe₂/MoSe₂/WSe₂ heterostructure.** (A). Bottom exciton potential (left panel) and electron potential in the middle MoSe₂ layer (right panel). The electron potential has three degenerate local minima in the unit cell (red dashed lines), which we consider as the first orbital.
(B) Filling examples for $n_e = 1$ and $n_e = 2$. Blue dots are sites occupied by electrons, and the grey dots are unoccupied sites. (C) Example of energy shifts for one dipolar exciton with electron doping. The energy shift distribution is narrower than the experimental one in the main text and effective length $r_{eff}$ is longer than experimental value 7-8 nm.

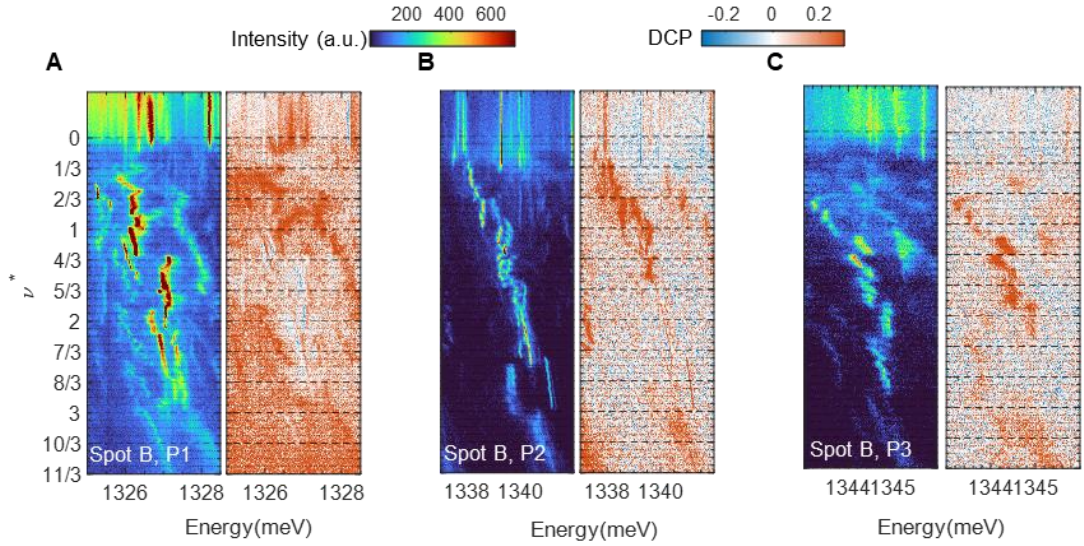

**Fig. S19. Polarization resolved spectra of other peaks at spot B.** PL intensity and DCP as a function of filling for peak P1 (A), peak P2 (B), and peak P3 (C). All the peaks show that localized interlayer excitons are unpolarized in the intrinsic region and become polarized when the electron doping starts, which is attributed to the trion formation. Thereafter, the excitons become unpolarized at certain filling range and finally recover the polarization at higher filling. The excitation laser is $\sigma^-$ polarized at 1.70 eV and 50 nW.

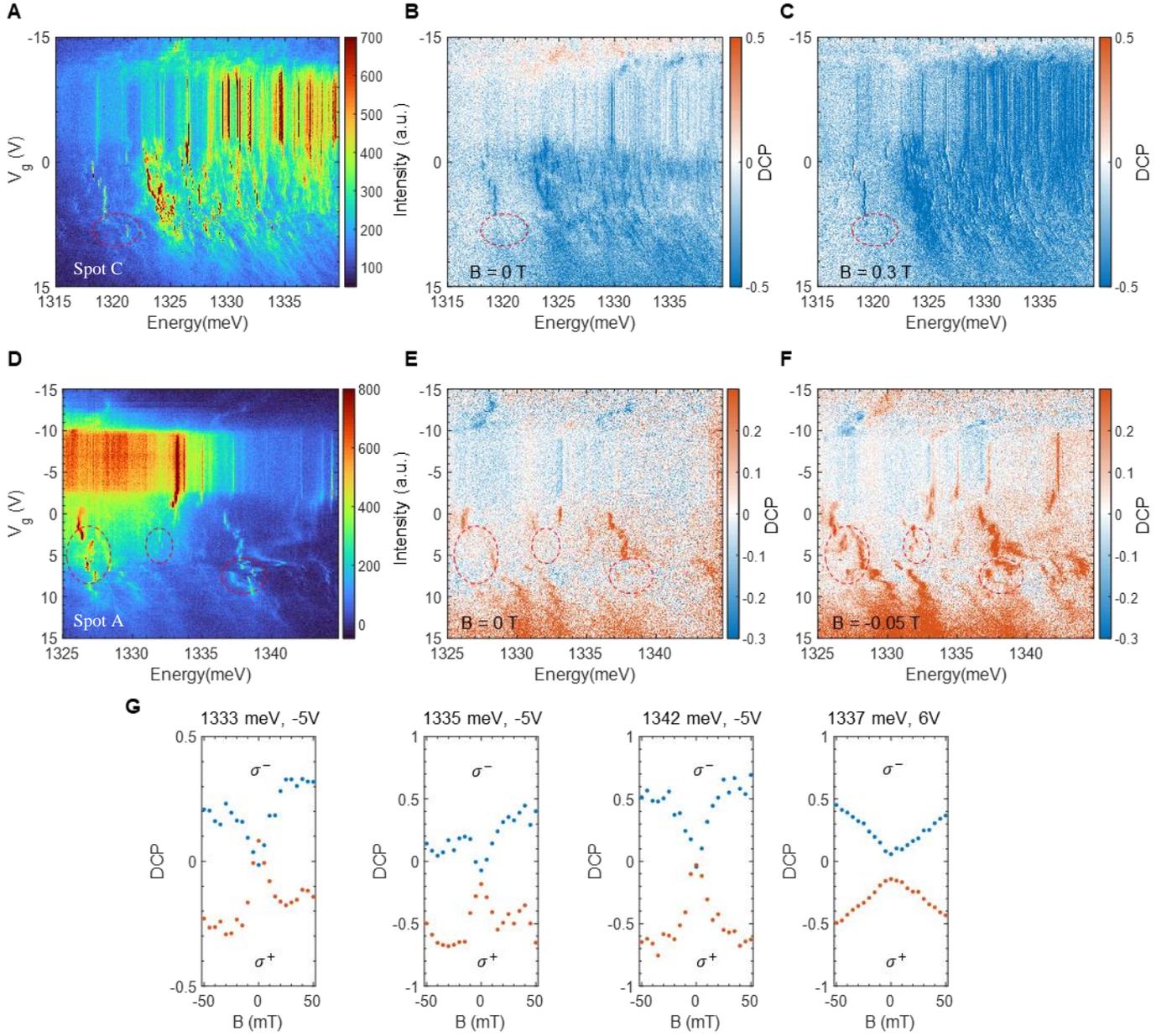

**Fig. S20. The recovery of circular polarization under small magnetic field in intrinsic and electron doped regions.** (A-C). PL intensity, DCP at $B$ = 0, and DCP at $B$ = 0.3 T at spot C. (D-F). PL intensity, DCP at $B$ = 0, and DCP at $B$ = -0.05 T at spot A. (G). Magnetic field dependence of DCP at -5 V (intrinsic) and 6 V (electron doped). The DCP dip is the signature of quenching electron-hole exchange interactions. The DCP at hole doping is opposite to that at electron doping. The red dashed circles are the ranges where the DCP disappears at 0 T but recovers at small finite $B$ field. The excitation laser is $\sigma^+$ polarized for spot C and at $\sigma^-$ polarized for spot A.

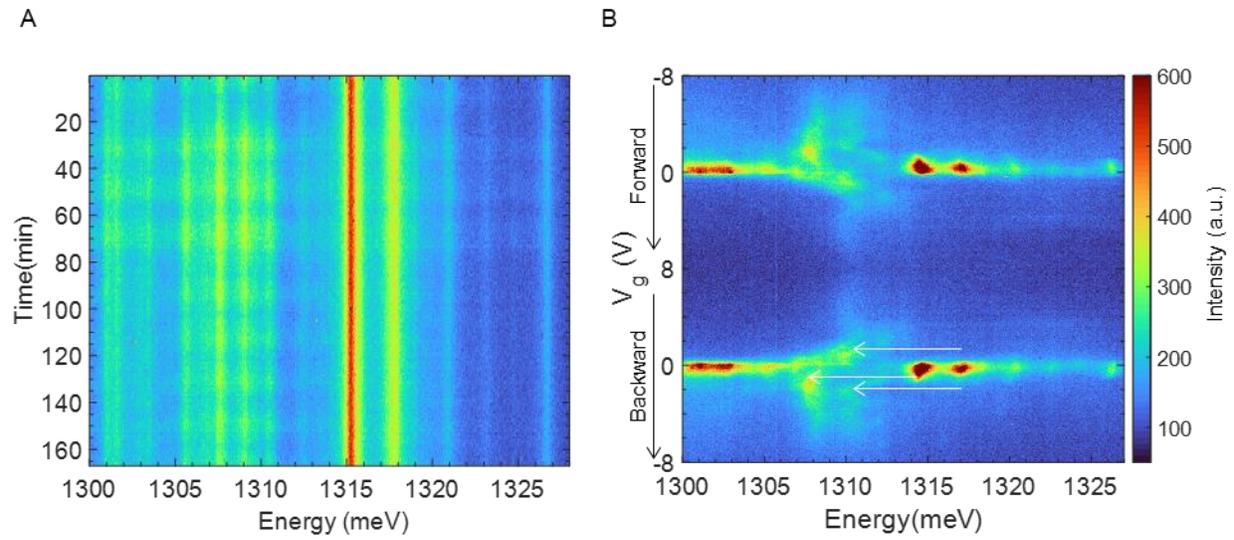

**Fig. S21. Time-trace and gate dependence of PL emission in device 2.** (A). Time trace shows no jittering of localized interlayer excitons. (B). Trion formation and reproducible red- and blue-shifts of localized interlayer excitons. The white arrows are 7 meV energy shifts. The linewidth is 500 µeV at 7 K.




\begin{thebibliography}{10}
\expandafter\ifx\csname url\endcsname\relax
  \def\url#1{\texttt{#1}}\fi
\expandafter\ifx\csname urlprefix\endcsname\relax\def\urlprefix{URL }\fi
\providecommand{\bibinfo}[2]{#2}
\providecommand{\eprint}[2][]{\url{#2}}

\bibitem{bistritzer2011moire}
\bibinfo{author}{Bistritzer, R.} \& \bibinfo{author}{MacDonald, A.~H.}
\newblock \bibinfo{title}{Moir{\'e} bands in twisted double-layer graphene}.
\newblock \emph{\bibinfo{journal}{Proceedings of the National Academy of
  Sciences}} \textbf{\bibinfo{volume}{108}}, \bibinfo{pages}{12233--12237}
  (\bibinfo{year}{2011}).

\bibitem{lisi2021observation}
\bibinfo{author}{Lisi, S.} \emph{et~al.}
\newblock \bibinfo{title}{Observation of flat bands in twisted bilayer
  graphene}.
\newblock \emph{\bibinfo{journal}{Nature Physics}}
  \textbf{\bibinfo{volume}{17}}, \bibinfo{pages}{189--193}
  (\bibinfo{year}{2021}).

\bibitem{regan2020mott}
\bibinfo{author}{Regan, E.~C.} \emph{et~al.}
\newblock \bibinfo{title}{Mott and generalized wigner crystal states in wse
  2/ws 2 moir{\'e} superlattices}.
\newblock \emph{\bibinfo{journal}{Nature}} \textbf{\bibinfo{volume}{579}},
  \bibinfo{pages}{359--363} (\bibinfo{year}{2020}).

\bibitem{zhou2021bilayer}
\bibinfo{author}{Zhou, Y.} \emph{et~al.}
\newblock \bibinfo{title}{Bilayer wigner crystals in a transition metal
  dichalcogenide heterostructure}.
\newblock \emph{\bibinfo{journal}{Nature}} \textbf{\bibinfo{volume}{595}},
  \bibinfo{pages}{48--52} (\bibinfo{year}{2021}).

\bibitem{smolenski2021signatures}
\bibinfo{author}{Smole{\'n}ski, T.} \emph{et~al.}
\newblock \bibinfo{title}{Signatures of wigner crystal of electrons in a
  monolayer semiconductor}.
\newblock \emph{\bibinfo{journal}{Nature}} \textbf{\bibinfo{volume}{595}},
  \bibinfo{pages}{53--57} (\bibinfo{year}{2021}).

\bibitem{shimazaki2021optical}
\bibinfo{author}{Shimazaki, Y.} \emph{et~al.}
\newblock \bibinfo{title}{Optical signatures of periodic charge distribution in
  a mott-like correlated insulator state}.
\newblock \emph{\bibinfo{journal}{Physical Review X}}
  \textbf{\bibinfo{volume}{11}}, \bibinfo{pages}{021027}
  (\bibinfo{year}{2021}).

\bibitem{li2021quantum}
\bibinfo{author}{Li, T.} \emph{et~al.}
\newblock \bibinfo{title}{Quantum anomalous hall effect from intertwined
  moir{\'e} bands}.
\newblock \emph{\bibinfo{journal}{arXiv preprint arXiv:2107.01796}}
  (\bibinfo{year}{2021}).

\bibitem{chen2020tunable}
\bibinfo{author}{Chen, G.} \emph{et~al.}
\newblock \bibinfo{title}{Tunable correlated chern insulator and ferromagnetism
  in a moir{\'e} superlattice}.
\newblock \emph{\bibinfo{journal}{Nature}} \textbf{\bibinfo{volume}{579}},
  \bibinfo{pages}{56--61} (\bibinfo{year}{2020}).

\bibitem{polshyn2020electrical}
\bibinfo{author}{Polshyn, H.} \emph{et~al.}
\newblock \bibinfo{title}{Electrical switching of magnetic order in an orbital
  chern insulator}.
\newblock \emph{\bibinfo{journal}{Nature}} \textbf{\bibinfo{volume}{588}},
  \bibinfo{pages}{66--70} (\bibinfo{year}{2020}).

\bibitem{kennes2021moire}
\bibinfo{author}{Kennes, D.~M.} \emph{et~al.}
\newblock \bibinfo{title}{Moir{\'e} heterostructures as a condensed-matter
  quantum simulator}.
\newblock \emph{\bibinfo{journal}{Nature Physics}}
  \textbf{\bibinfo{volume}{17}}, \bibinfo{pages}{155--163}
  (\bibinfo{year}{2021}).

\bibitem{allain2015electrical}
\bibinfo{author}{Allain, A.}, \bibinfo{author}{Kang, J.},
  \bibinfo{author}{Banerjee, K.} \& \bibinfo{author}{Kis, A.}
\newblock \bibinfo{title}{Electrical contacts to two-dimensional
  semiconductors}.
\newblock \emph{\bibinfo{journal}{Nature materials}}
  \textbf{\bibinfo{volume}{14}}, \bibinfo{pages}{1195--1205}
  (\bibinfo{year}{2015}).

\bibitem{li2021charge}
\bibinfo{author}{Li, T.} \emph{et~al.}
\newblock \bibinfo{title}{Charge-order-enhanced capacitance in semiconductor
  moiré superlattices}.
\newblock \emph{\bibinfo{journal}{Nature Nanotechnology}}
  \textbf{\bibinfo{volume}{16}}, \bibinfo{pages}{1068--1072}
  (\bibinfo{year}{2021}).

\bibitem{tang2020simulation}
\bibinfo{author}{Tang, Y.} \emph{et~al.}
\newblock \bibinfo{title}{Simulation of hubbard model physics in wse 2/ws 2
  moir{\'e} superlattices}.
\newblock \emph{\bibinfo{journal}{Nature}} \textbf{\bibinfo{volume}{579}},
  \bibinfo{pages}{353--358} (\bibinfo{year}{2020}).

\bibitem{xu2020correlated}
\bibinfo{author}{Xu, Y.} \emph{et~al.}
\newblock \bibinfo{title}{Correlated insulating states at fractional fillings
  of moir{\'e} superlattices}.
\newblock \emph{\bibinfo{journal}{Nature}} \textbf{\bibinfo{volume}{587}},
  \bibinfo{pages}{214--218} (\bibinfo{year}{2020}).

\bibitem{jin2021stripe}
\bibinfo{author}{Jin, C.} \emph{et~al.}
\newblock \bibinfo{title}{Stripe phases in wse 2/ws 2 moir{\'e} superlattices}.
\newblock \emph{\bibinfo{journal}{Nature Materials}} \bibinfo{pages}{1--5}
  (\bibinfo{year}{2021}).

\bibitem{liu2021excitonic}
\bibinfo{author}{Liu, E.} \emph{et~al.}
\newblock \bibinfo{title}{Excitonic and valley-polarization signatures of
  fractional correlated electronic phases in a wse 2/ws 2 moir{\'e}
  superlattice}.
\newblock \emph{\bibinfo{journal}{Physical Review Letters}}
  \textbf{\bibinfo{volume}{127}}, \bibinfo{pages}{037402}
  (\bibinfo{year}{2021}).

\bibitem{huang2021correlated}
\bibinfo{author}{Huang, X.} \emph{et~al.}
\newblock \bibinfo{title}{Correlated insulating states at fractional fillings
  of the ws2/wse2 moir{\'e} lattice}.
\newblock \emph{\bibinfo{journal}{Nature Physics}}
  \textbf{\bibinfo{volume}{17}}, \bibinfo{pages}{715--719}
  (\bibinfo{year}{2021}).

\bibitem{li2021imaging}
\bibinfo{author}{Li, H.} \emph{et~al.}
\newblock \bibinfo{title}{Imaging two-dimensional generalized wigner crystals}.
\newblock \emph{\bibinfo{journal}{Nature}} \textbf{\bibinfo{volume}{597}},
  \bibinfo{pages}{650--654} (\bibinfo{year}{2021}).

\bibitem{tong2020interferences}
\bibinfo{author}{Tong, Q.}, \bibinfo{author}{Chen, M.}, \bibinfo{author}{Xiao,
  F.}, \bibinfo{author}{Yu, H.} \& \bibinfo{author}{Yao, W.}
\newblock \bibinfo{title}{Interferences of electrostatic moir{\'e} potentials
  and bichromatic superlattices of electrons and excitons in transition metal
  dichalcogenides}.
\newblock \emph{\bibinfo{journal}{2D Materials}} \textbf{\bibinfo{volume}{8}},
  \bibinfo{pages}{025007} (\bibinfo{year}{2020}).

\bibitem{florens2002mott}
\bibinfo{author}{Florens, S.}, \bibinfo{author}{Georges, A.},
  \bibinfo{author}{Kotliar, G.} \& \bibinfo{author}{Parcollet, O.}
\newblock \bibinfo{title}{Mott transition at large orbital degeneracy:
  Dynamical mean-field theory}.
\newblock \emph{\bibinfo{journal}{Physical Review B}}
  \textbf{\bibinfo{volume}{66}}, \bibinfo{pages}{205102}
  (\bibinfo{year}{2002}).

\bibitem{zhang2021electronic}
\bibinfo{author}{Zhang, Y.}, \bibinfo{author}{Liu, T.} \& \bibinfo{author}{Fu,
  L.}
\newblock \bibinfo{title}{Electronic structures, charge transfer, and charge
  order in twisted transition metal dichalcogenide bilayers}.
\newblock \emph{\bibinfo{journal}{Physical Review B}}
  \textbf{\bibinfo{volume}{103}}, \bibinfo{pages}{155142}
  (\bibinfo{year}{2021}).

\bibitem{methods}
 \textbf{\bibinfo{volume}{See Methods section}}.

\bibitem{seyler2019signatures}
\bibinfo{author}{Seyler, K.~L.} \emph{et~al.}
\newblock \bibinfo{title}{Signatures of moir{\'e}-trapped valley excitons in
  mose 2/wse 2 heterobilayers}.
\newblock \emph{\bibinfo{journal}{Nature}} \textbf{\bibinfo{volume}{567}},
  \bibinfo{pages}{66--70} (\bibinfo{year}{2019}).

\bibitem{wang2021moire}
\bibinfo{author}{Wang, X.} \emph{et~al.}
\newblock \bibinfo{title}{Moir{\'e} trions in mose2/wse2 heterobilayers}.
\newblock \emph{\bibinfo{journal}{Nature Nanotechnology}} \bibinfo{pages}{1--6}
  (\bibinfo{year}{2021}).

\bibitem{liu2021signatures}
\bibinfo{author}{Liu, E.} \emph{et~al.}
\newblock \bibinfo{title}{Signatures of moir{\'e} trions in wse2/mose2
  heterobilayers}.
\newblock \emph{\bibinfo{journal}{Nature}} \textbf{\bibinfo{volume}{594}},
  \bibinfo{pages}{46--50} (\bibinfo{year}{2021}).

\bibitem{PhysRevX.11.031033}
\bibinfo{author}{Brotons-Gisbert, M.} \emph{et~al.}
\newblock \bibinfo{title}{Moir\'e-trapped interlayer trions in a charge-tunable
  ${\mathrm{wse}}_{2}/{\mathrm{mose}}_{2}$ heterobilayer}.
\newblock \emph{\bibinfo{journal}{Phys. Rev. X}} \textbf{\bibinfo{volume}{11}},
  \bibinfo{pages}{031033} (\bibinfo{year}{2021}).

\bibitem{baek2021coulombstaircase}
\bibinfo{author}{Baek, H.} \emph{et~al.}
\newblock \bibinfo{title}{Optical read-out of coulomb staircases in a moir{\'e}
  superlattice via trapped interlayer trions}.
\newblock \emph{\bibinfo{journal}{Nature Nanotechnology}} \bibinfo{pages}{1--7}
  (\bibinfo{year}{2021}).

\bibitem{yu2014dirac}
\bibinfo{author}{Yu, H.}, \bibinfo{author}{Liu, G.-B.}, \bibinfo{author}{Gong,
  P.}, \bibinfo{author}{Xu, X.} \& \bibinfo{author}{Yao, W.}
\newblock \bibinfo{title}{Dirac cones and dirac saddle points of bright
  excitons in monolayer transition metal dichalcogenides}.
\newblock \emph{\bibinfo{journal}{Nature communications}}
  \textbf{\bibinfo{volume}{5}}, \bibinfo{pages}{1--7} (\bibinfo{year}{2014}).

\bibitem{lu2019optical}
\bibinfo{author}{Lu, X.} \emph{et~al.}
\newblock \bibinfo{title}{Optical initialization of a single spin-valley in
  charged wse 2 quantum dots}.
\newblock \emph{\bibinfo{journal}{Nature nanotechnology}}
  \textbf{\bibinfo{volume}{14}}, \bibinfo{pages}{426--431}
  (\bibinfo{year}{2019}).

\bibitem{smolenski2016tuning}
\bibinfo{author}{Smole{\'n}ski, T.} \emph{et~al.}
\newblock \bibinfo{title}{Tuning valley polarization in a wse 2 monolayer with
  a tiny magnetic field}.
\newblock \emph{\bibinfo{journal}{Physical Review X}}
  \textbf{\bibinfo{volume}{6}}, \bibinfo{pages}{021024} (\bibinfo{year}{2016}).

\bibitem{georges2013strong}
\bibinfo{author}{Georges, A.}, \bibinfo{author}{Medici, L.~d.} \&
  \bibinfo{author}{Mravlje, J.}
\newblock \bibinfo{title}{Strong correlations from hund’s coupling}.
\newblock \emph{\bibinfo{journal}{Annu. Rev. Condens. Matter Phys.}}
  \textbf{\bibinfo{volume}{4}}, \bibinfo{pages}{137--178}
  (\bibinfo{year}{2013}).

\bibitem{zomer2014fast}
\bibinfo{author}{Zomer, P.}, \bibinfo{author}{Guimar{\~a}es, M.},
  \bibinfo{author}{Brant, J.}, \bibinfo{author}{Tombros, N.} \&
  \bibinfo{author}{Van~Wees, B.}
\newblock \bibinfo{title}{Fast pick up technique for high quality
  heterostructures of bilayer graphene and hexagonal boron nitride}.
\newblock \emph{\bibinfo{journal}{Applied Physics Letters}}
  \textbf{\bibinfo{volume}{105}}, \bibinfo{pages}{013101}
  (\bibinfo{year}{2014}).

\bibitem{kim2016van}
\bibinfo{author}{Kim, K.} \emph{et~al.}
\newblock \bibinfo{title}{van der waals heterostructures with high accuracy
  rotational alignment}.
\newblock \emph{\bibinfo{journal}{Nano letters}} \textbf{\bibinfo{volume}{16}},
  \bibinfo{pages}{1989--1995} (\bibinfo{year}{2016}).

\bibitem{kresse1996efficient}
\bibinfo{author}{Kresse, G.} \& \bibinfo{author}{Furthm{\"u}ller, J.}
\newblock \bibinfo{title}{Efficient iterative schemes for ab initio
  total-energy calculations using a plane-wave basis set}.
\newblock \emph{\bibinfo{journal}{Physical review B}}
  \textbf{\bibinfo{volume}{54}}, \bibinfo{pages}{11169} (\bibinfo{year}{1996}).

\bibitem{perdew1996generalized}
\bibinfo{author}{Perdew, J.~P.}, \bibinfo{author}{Burke, K.} \&
  \bibinfo{author}{Ernzerhof, M.}
\newblock \bibinfo{title}{Generalized gradient approximation made simple}.
\newblock \emph{\bibinfo{journal}{Physical review letters}}
  \textbf{\bibinfo{volume}{77}}, \bibinfo{pages}{3865} (\bibinfo{year}{1996}).

\bibitem{klimevs2011van}
\bibinfo{author}{Klime{\v{s}}, J.}, \bibinfo{author}{Bowler, D.~R.} \&
  \bibinfo{author}{Michaelides, A.}
\newblock \bibinfo{title}{Van der waals density functionals applied to solids}.
\newblock \emph{\bibinfo{journal}{Physical Review B}}
  \textbf{\bibinfo{volume}{83}}, \bibinfo{pages}{195131}
  (\bibinfo{year}{2011}).

\bibitem{wang2021vaspkit}
\bibinfo{author}{Wang, V.}, \bibinfo{author}{Xu, N.}, \bibinfo{author}{Liu,
  J.-C.}, \bibinfo{author}{Tang, G.} \& \bibinfo{author}{Geng, W.-T.}
\newblock \bibinfo{title}{Vaspkit: a user-friendly interface facilitating
  high-throughput computing and analysis using vasp code}.
\newblock \emph{\bibinfo{journal}{Computer Physics Communications}}
  \bibinfo{pages}{108033} (\bibinfo{year}{2021}).

\end{thebibliography}
\end{document}